\documentclass[structabstract]{aa}  
\usepackage{graphicx}
\usepackage{color}  
\usepackage{amsmath}
\usepackage{txfonts}
\usepackage{natbib}
\usepackage{subfigure}
\begin{document}
\graphicspath{{images/}}
   \title{A coordinated optical and X-ray spectroscopic campaign\\ on \object{HD\,179949}: searching for planet-induced\\ chromospheric and coronal activity.}


   \author{G.\ Scandariato\inst{1}
          \and
          A.\ Maggio\inst{2}
          \and
          A.F.\ Lanza\inst{1}
          \and
          I.\ Pagano\inst{1}
          \and
          R.\ Fares\inst{3}
          \and
          E.L.\ Shkolnik\inst{4}
          \and
          D.\ Bohlender\inst{5}
          \and
          A.C.\ Cameron\inst{3}
          \and
          S.\ Dieters\inst{6}
          \and
          J.-F.\ Donati\inst{7}
          \and
          A.F.\ Mart\'inez Fiorenzano\inst{8}
          \and
          M.\ Jardine\inst{3}
          \and
          C.\ Moutou\inst{9}
          }

   \institute{INAF Osservatorio Astrofisico di Catania, via S.\ Sofia 78, 95123 Catania, Italy\\
         \and
             INAF Osservatorio Astronomico di Palermo "G.~S.~Vaiana", Piazza del Parlamento 1, 90134 Palermo, Italy\\
         \and
             School of Physics and Astronomy, Univ.\ of St Andrews, St Andrews, Scotland KY16 9SS, UK\\
          	\and
            Lowell Observatory, 1400 W.\ Mars Hill Road, Flagstaff, AZ, 86001 USA\\
            \and
            	HIA/NRC, 5071 West Saanich Road, Victoria, BC V9E 2E7, Canada\\
             \and
            	School of Mathematics and Physics, University of Tasmania, PB 37 GP0 Hobart, Tasmania 7001, Australia\\
          \and
			IRAP-UMR 5277, CNRS \& Univ.\ de Toulouse, 14 Av.\ E.\ Belin, F-31400 Toulouse, France\\
            \and
            	Fundaci\'on Galileo Galilei - INAF, Rambla Jos\'e Ana Fern\'andez P\'erez, 7, 38712 Bre\~na Baja, TF - Spain\\
          \and
          	LAM-UMR 6110, CNRS \& Univ.\ de Provence, 38 rue Fr\'ederic Juliot-Curie, F-13013 Marseille, France\\
				\mbox{}\\
              \email{gas@oact.inaf.it}\\
            }


\abstract
{\object{HD\,179949} is an F8V star, orbited by a close-in giant planet with a period of $\sim 3$~days. Previous studies suggested that the planet enhances the magnetic activity of the parent star, producing a chromospheric hot spot which rotates in phase with the planet orbit. However, this phenomenon is intermittent since it was observed in several but not all seasons.}
{A long-term monitoring of the magnetic activity of \object{HD\,179949} is required to study the amplitude and time scales of star-planet interactions.}
{In 2009 we performed a simultaneous optical and X-ray spectroscopic campaign to monitor the magnetic activity of \object{HD\,179949} during $\sim$5 orbital periods and $\sim$2 stellar rotations. We analyzed the \ion{Ca}{ii} H\&K lines as a proxy for chromospheric activity, and we studied the X-ray emission in search of flux modulations and to determine basic properties of the coronal plasma.}
{A detailed analysis of the flux in the cores of the \ion{Ca}{ii} H\&K lines and a similar study of the X-ray photometry shows evidence of source variability, including one flare. The analysis of the the time series of chromospheric data indicates a modulation with a $\sim$11~days period, compatible with the stellar rotation period at high latitudes. Instead, the X-ray light curve suggests a signal with a period of $\sim$4~days, consistent with the presence of two active regions on opposite hemispheres.}
{The observed variability can be explained, most likely, as due to rotational modulation and to intrinsic evolution of chromospheric and coronal activity. There is no clear signature related to the orbital motion of the planet, but the possibility that just a fraction of the chromospheric and coronal variability is modulated with the orbital period of the planet, or the stellar-planet beat period, cannot be excluded. We conclude that any effect due to the presence of the planet is difficult to disentangle.} 


	\titlerunning{A coordinated spectroscopic campaign on HD\,179949}
	\authorrunning{Scandariato et al.}

   \maketitle
%

\section{Introduction}

Approximately 20\% of the planets detected so far are giants ($M_p \sin i > 0.2\,M_J$)  in tight orbits around their parent stars (semimajor axis $a < 0.1$~AU). These planets are dubbed Hot Jupiters (HJs). Star-Planet tidal and/or magnetospheric Interaction (SPI) in these systems were claimed to be the possible cause of stellar activity enhancements  \citep{2000ApJ...529.1031R, 2000ApJ...533L.151C}. A diagnostic of SPI is a signature of stellar activity observed to vary with the planet orbital motion, e.g., the Ca II H\&K lines, the X-ray emission or the optical continuum, which sample the chromospheric, coronal and photospheric levels of the atmosphere, respectively. \citet{2000ApJ...533L.151C} suggested that  in the case of tidal interactions, activity is modulated  by half of the orbital period, while in the case of magnetospheric interactions it varies with the orbital period. Recently, \citet{2010MNRAS.406..409F} suggested that the enhancement due to magnetospheric SPIs is more likely modulated with the beat period of the system, i.e.\ the synodic period between the stellar rotation and the orbital period.

 \citet{2003ApJ...597.1092S,2005ApJ...622.1075S,2008ApJ...676..628S} found the chromospheric emission of a few stars hosting  HJs to be variable with the planet orbital period. Models assuming a magnetic interaction similar to that in the Jupiter-Io system fail to account for  the observed chromospheric excess powers \citep[cf.][and references therein]{2009A&A...505..339L}. Other authors suggest that  a magnetic field reconfiguration on a  scale comparable with the star-planet separation can be triggered by  localized reconnection at the boundary between the stellar corona and the planetary magnetosphere  \citep[cf., e.g., ][]{2009A&A...505..339L,2011Ap&SS.336..303L,2009ApJ...704L..85C}. Because of the large coronal volumes  involved in the energy release, these star-planet magnetic interaction (SPMI) models succeed in accounting for  the observed energy budget.

Space-based optical photometry has also provided some evidence of SPI. The variance of CoRoT-2 stellar flux was found to be modulated in phase with the planet orbital period, reaching a maximum near the time of transit \citep{2009EM&P..105..373P}.  The Microvariability and Oscillation of STars (MOST) space telescope found a spot that persisted for hundreds of rotation cycles on $\tau$\,Boo \citep{2008A&A...482..691W}, while \citet{2009A&A...506..255L, 2011A&A...525A..14L} reported hints for SPI  in CoRoT-4 and CoRoT-6. 

A statistical survey of the X-ray emission of stars with close-in HJs ($a < 0.15$~AU) suggested that they may be $\approx 2-4$ times more active than stars with more distant planets ($a > 1.5 $~AU; Kashyap et al. 2008). On the other hand, \citet{2010A&A...515A..98P} and \citet{2011AN....332.1052P} suggested that the coronae of cool stars exhibit a considerable amount of intrinsic variability compared to the effects expected from SPI, and that the observations  of emission enhancements attributed to SPI can be explained by selection effects. Nonetheless, magnetohydrodynamic (MHD) simulations by \citet{2009ApJ...704L..85C} show that SPI may increase the X-ray luminosity. 

Observational studies of HJ hosting stars show that not all systems display hints of interaction. Moreover, individual systems may show seasons of activity enhancement and seasons without such effects. Some systems show a phase lag between the subplanetary longitude and the peak of the activity enhancement \citep{2003ApJ...597.1092S,2005ApJ...622.1075S,2008ApJ...676..628S}. Different models try to account for these phase lags, either considering the case where the stellar field is a tilted dipole \citep{2006MNRAS.367L...1M}, or adopting a model based on the propagation of Alfv\`en waves along the sub-Alfv\`enic stellar wind flow relative to the planet \citep{2006A&A...460..317P}, or considering a non-potential magnetic field configuration for the closed coronal field \citep{2008A&A...487.1163L}. Finally, \citet{2007astro.ph..2530C} modeled the \ion{Ca}{ii} H\&K light curve of a HJ hosting star for different configurations of the stellar magnetic field based on solar observations, and showed that the SPI may be triggered or may not, depending on the specific magnetic field geometry. 

\object{HD\,179949} ($d \simeq 27$~pc) is an F8V star, orbited by a giant planet with mass $ M\sin  i \geq 0.916 \pm 0.076$~M$_{J}$, orbital semi-major axis $a =0.0443 \pm0.026$~AU, and period of 3.092~days \citep{2006ApJ...646..505B}. The star showed several seasons of activity enhancement modulated with the orbital period, with a chromospheric hot spot leading the planet by $\sim 70^{\circ}-80^{\circ}$. Seasons with no apparent magnetic interaction between the star and the planet \citep{2008ApJ...676..628S,Gurdemiretal12} were also observed. 

We organized a multi-wavelength campaign to observe \object{HD\,179949} in September and October 2009, and to obtain nearly simultaneous optical and X-ray photometry and spectroscopy. Our aim was a quantitative characterization of the stellar activity to develop models of the star-planet interaction. We remark that data series limited to a fraction of the stellar rotation or planet orbital period may not provide sufficient information to infer the correct time scale and origin of the observed phenomena \citep{Miller2012}; this is not the case of the present study, since it is based on optical observations which covered a time range of $\sim 16$~days, i.e.\ about two stellar rotations and five orbital periods,
and X-ray observations spanning more than one month.

Spectropolarimetric data acquired during our observation campaign were already presented by \citet{Faresetal12}. In the present paper, we focus our attention on the temporal evolution of chromospheric and coronal activity, with the analysis of the \ion{Ca}{ii} H\&K lines ($\lambda\lambda$ 3968.487, 3933.673\,\AA) (Sect.~2), and the analysis of X-ray data (Sect.~\ref{sec:xdata}) obtained in the framework of an XMM-Newton Large Program. The main results and conclusions are reported in Sect.~4.


\section{Optical spectroscopy}

\subsection{Observations and data reduction}\label{sec:data}

\subsubsection{ESPaDOnS spectra}\label{sec:espadons}
ESPaDOnS is a high-resolution spectropolarimeter installed at the 3.6-m Canada-France-Hawaii Telescope (CFHT) on Mauna Kea. It provides spectra spanning the whole optical domain (3700 to 10000~\AA) with a resolution of R$\sim$65,000 (when used in spectropolarimetric mode). We typically collected two spectra per night, for a total exposure time of 9800~s (whenever possible) on each of 12 nights, between September 24 and October 9 2009. The data were reduced using the fully automated tool Libre-ESpRIT \citep{Donati97}.

\begin{figure*}
\centering
\includegraphics[width=.4\linewidth]{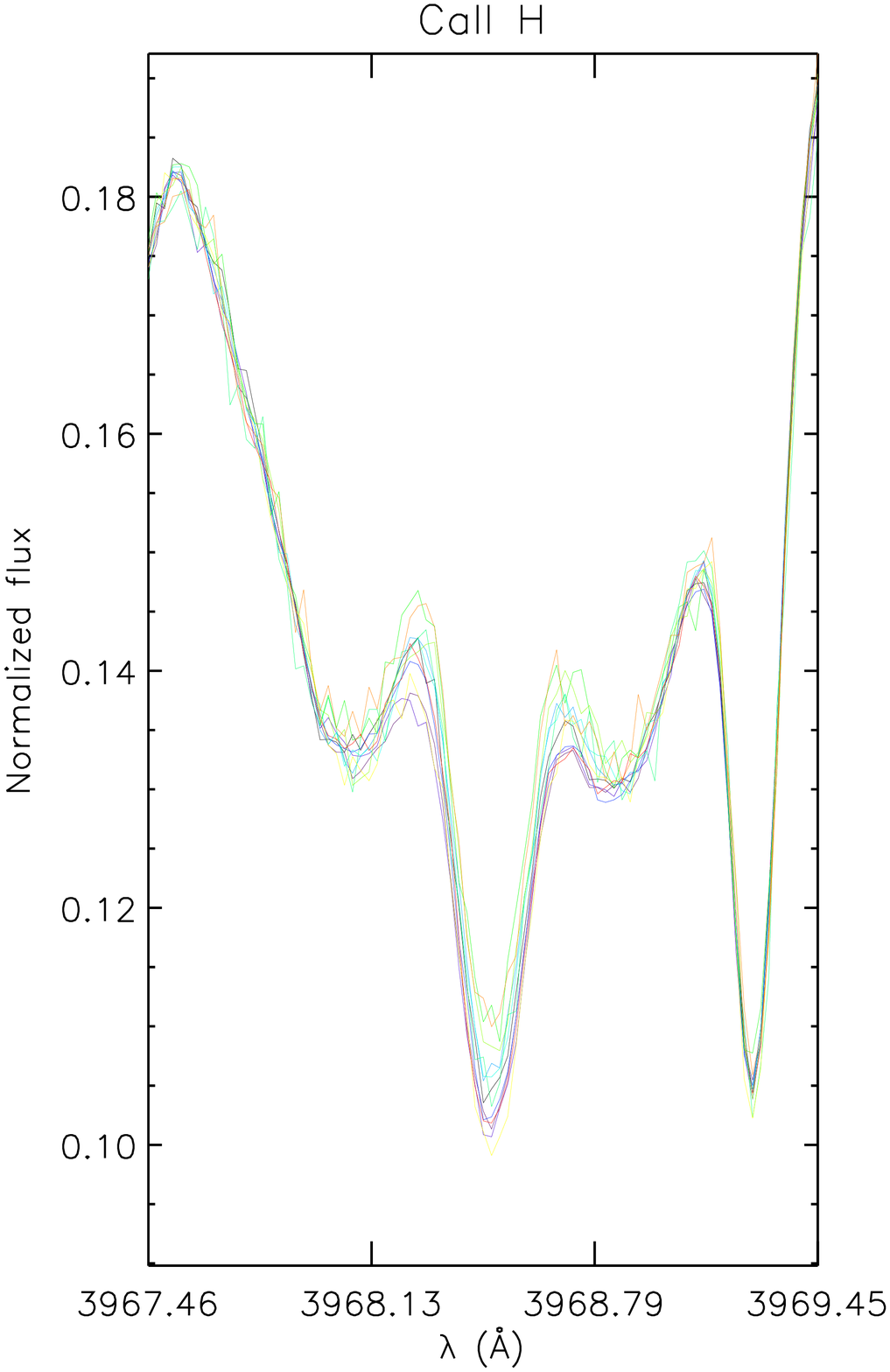}
\includegraphics[width=.4\linewidth]{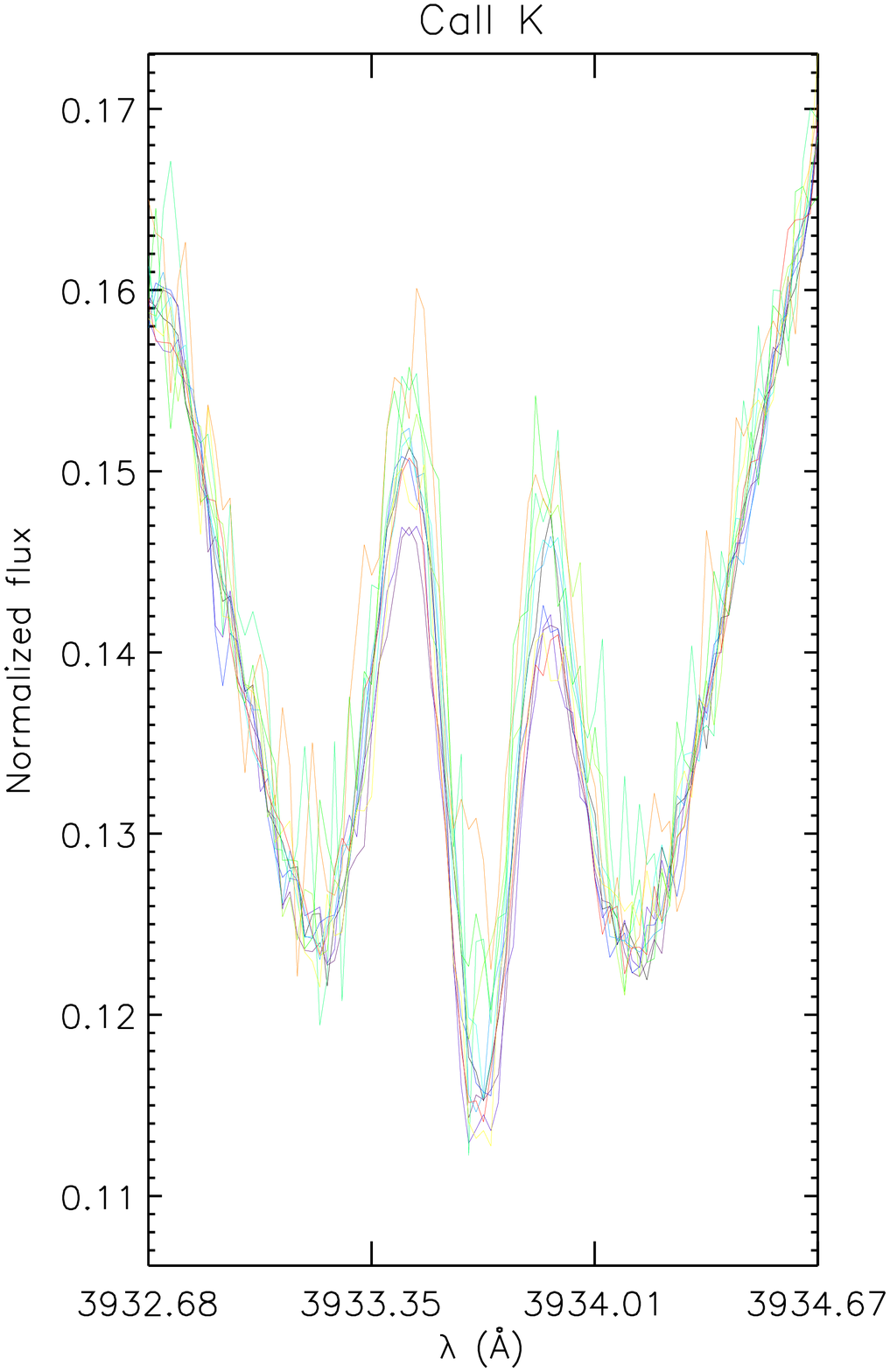}
\caption{Zoom-in of the cores of the \ion{Ca}{ii} H and K spectra (left and right panel respectively) of \object{HD\,179949} taken with ESPaDOns. Different colors code the nights of observation.}\label{fig:espadons}
\end{figure*}

In order to derive a common wavelength calibration, we choose the highest S/R spectrum as the reference spectrum and cross-correlate the other spectra in the series with this reference spectrum. 
When multiple observations per night are available, we compute the average spectrum, thus obtaining a master series of 12 spectra. The \ion{Ca}{ii} H and K spectra taken with ESPaDOnS are shown in Fig.~\ref{fig:espadons}, and their signal-to-noise ratio is reported in the observation log (Table~\ref{tab:log_espadons}).

\begin{figure*}
\centering
\includegraphics[width=.49\linewidth]{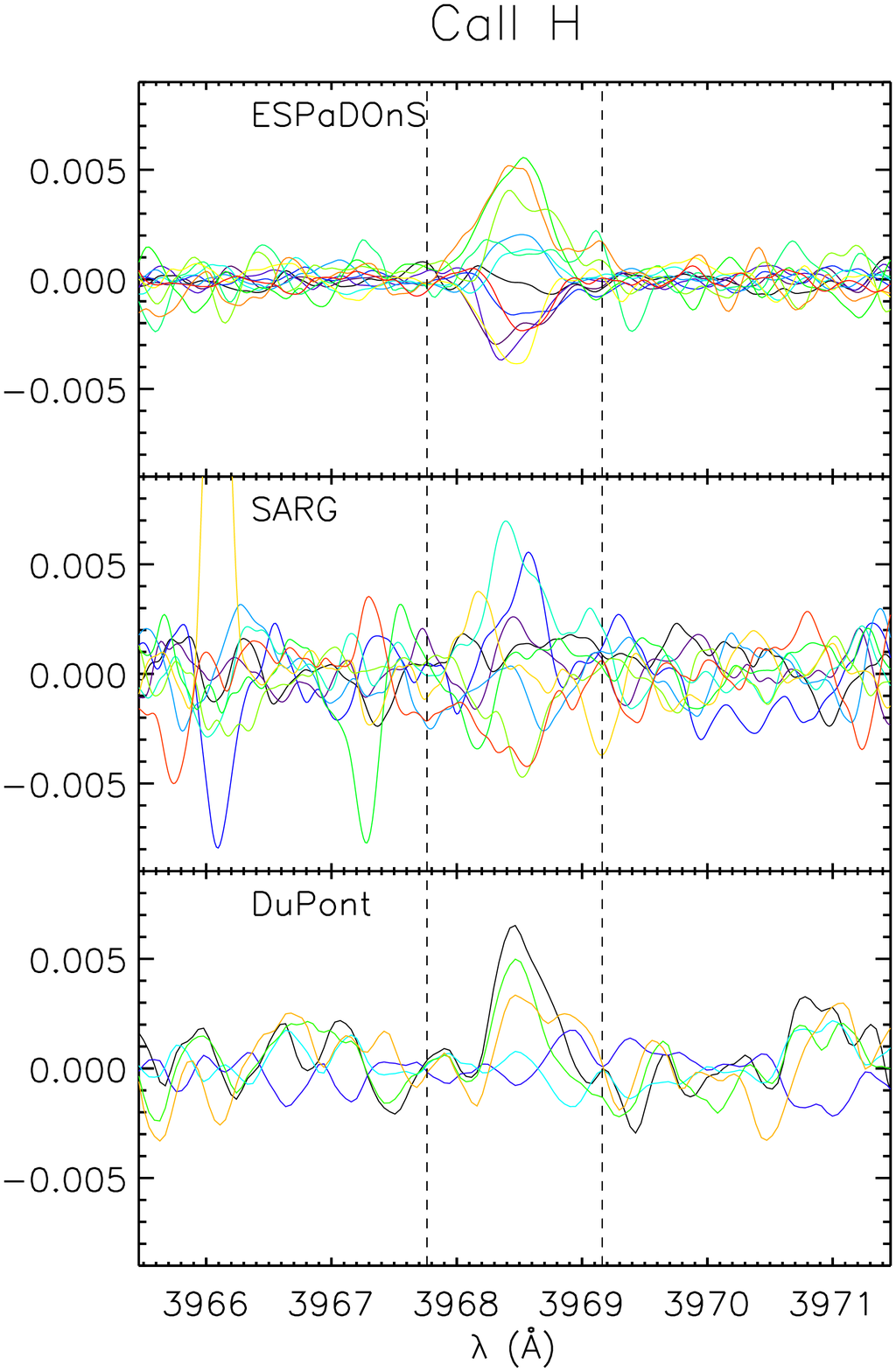}
\includegraphics[width=.49\linewidth]{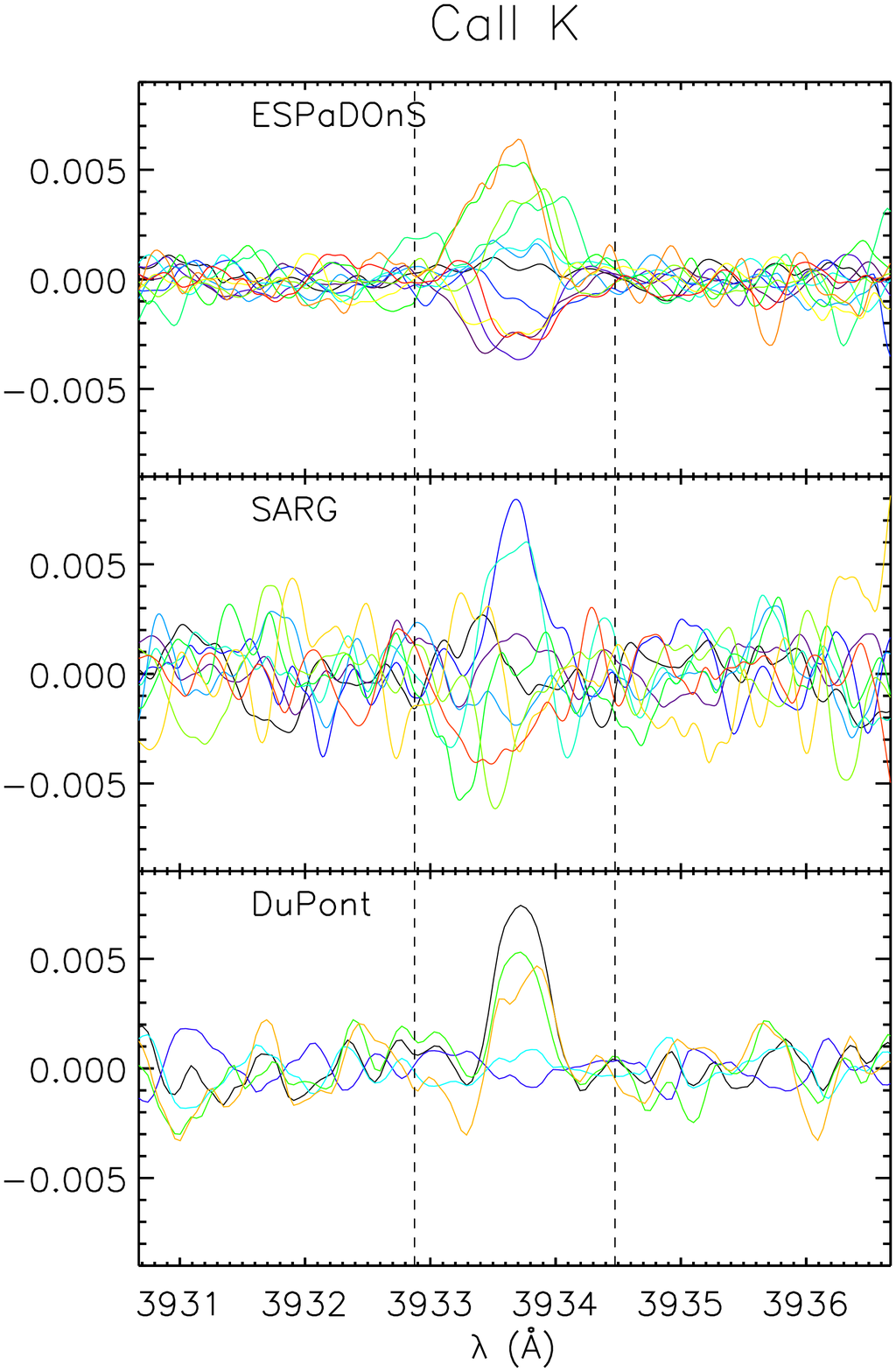}
\caption{Deviations from the \ion{Ca}{ii} H (left column) and K (right column) average spectrum, for the ESPaDOnS, SARG and du Pont series of spectra (from top to bottom, respectively). In each panel, colors code the different nights of observations. The vertical dashed lines delimit the core of the lines, as assumed in the present analysis (see the text).}\label{fig:residues}
\end{figure*}

\begin{table}
 \begin{center}
\caption{Log of ESPaDOnS observations, and S/N per pixel of the \ion{Ca}{ii} H\&K and the reference \ion{Fe}{i} $\lambda$3928.0827 line cores.}\label{tab:log_espadons}
\scriptsize
  \begin{tabular}{cccccc}
 \hline\hline
Date at mid exposure & Orbital & Exp. time & S/N$_{\rm H}$ & S/N$_{\rm K}$ & S/N$_{\rm FeI}$\\
(MJD-55095) & phase & (s) & & &\\
\hline
4.739 & 0.210 & 9800 & 264 & 141 & 133\\
6.727 & 0.854 & 9800 & 276 & 159 & 133\\
7.732 & 0.177 & 9800 & 242 & 141 & 118\\
8.731 & 0.500 & 9800 & 315 & 120 & 110\\
9.736 & 0.825 & 9800 & 241 & 140 & 126\\
10.735 & 0.148 & 9800 & 229 & 140 & 126\\
11.750 & 0.477 & 5040 & 108 & 59 & 54\\
12.717 & 0.789 & 2520 & 144 & 88 & 73\\
13.804 & 0.141 & 5040 & 157 & 110 & 94\\
14.731 & 0.440 & 9800 & 174 & 100 & 90\\
16.752 & 0.094 & 9800 & 127 & 69 & 58\\
19.734 & 0.059 & 9800 & 272 & 141 & 124\\
\hline
  \end{tabular}
 \end{center}
\end{table}
 
\subsubsection{SARG spectra}\label{sec:sarg}
SARG is a cross-dispersed echelle spectrograph installed at the Italian 3.6-m Telescopio Nazionale Galileo (TNG) in La Palma, Canary Islands. We collected a total of 18 spectra, covering the wavelength range 3600--5140~\AA~with a resolution power R=86,000 and an exposure time of 2100~s. Due to weather variability, we performed our observations on September 21, from October 1 to 3, and from October 6 to 10 2009 (9 nights), collecting two spectra per night.

We reduce the spectra using custom IRAF scripts. We performe the standard steps in extracting the echelle spectra, carefully defining a new set of echelle apertures and new wavelength calibrations for each night, thus avoiding the effects of instrumental instabilities. We also automatically remove evident spikes (due to either cosmic ray hits or hot pixels) in the extracted spectra by means of a custom IDL routine. Wavelength calibration is performed as discussed in Sect.~\ref{sec:espadons}.

Since the H and K lines cover most of the corresponding echelle order, no continuum extraction is possible. Thus, we normalize these spectra using the continuum extracted from ESPaDOnS spectra.

For each night of observations, we average the two collected spectra, obtaining a master series of 9 spectra, unevenly distributed between 21st September and 10th October 2009. The log of the observations is reported in Table~\ref{tab:log_sarg}.

\begin{table}
 \begin{center}
\caption{Log of SARG observations, and S/N per pixel of the \ion{Ca}{ii} H\&K and of the reference \ion{Fe}{i}  $\lambda$3928.0827 line cores.}\label{tab:log_sarg}
\scriptsize
  \begin{tabular}{cccccc}
 \hline\hline
Date at mid exposure & Orbital & Exp. time & S/N$_{\rm H}$ & S/N$_{\rm K}$ & S/N$_{\rm FeI}$\\
(MJD-55095) & phase & (s) & & & \\
\hline
11.320 & 0.338 & 4200 & 85 & 67 & 59\\
12.318 & 0.660 & 4200 & 76 & 41 & 37\\
13.318 & 0.984 & 4200 & 42 & 44 & 39\\
16.302 & 0.948 & 4200 & 66 & 50 & 46\\
17.305 & 0.273 & 4200 & 54 & 51 & 44\\
18.332 & 0.605 & 4200 & 55 & 43 & 39\\
19.343 & 0.932 & 4200 & 78 & 51 & 47\\
20.314 & 0.246 & 4200 & 43 & 28 & 25\\
\hline
  \end{tabular}
 \end{center}
\end{table}

\subsubsection{du Pont spectra}

We obtained additional high-resolution optical spectra of \object{HD\,179949} using the Echelle spectrograph mounted on the 2.5-m du Pont telescope at Las Campanas Observatory.  The instrument uses a TEK\#5 CCD camera with a 2k$\times$2k format and 24 micron pixels. Simultaneous wavelength coverage extends from 3700 to 9300 \AA\ with a spectral resolution of about 30,000 with a 1\arcsec slit.

Each stellar exposure is bias-subtracted and flat-fielded for pixel-to-pixel sensitivity variations. After optimal extraction, the one-dimensional spectra are wavelength calibrated with a Th/Ar arc. Finally the spectra are divided by a flat-field response to correct for each order's blaze function. 

Our observing run spans 5 consecutive nights from October 5 to 9 2009, collecting 10 to 50 spectra each night with exposure times of $\sim$100~s. We perform wavelength calibration and continuum normalization as described in Sect.~\ref{sec:espadons} and Sect.~\ref{sec:sarg}, respectively. For each night, we compute the average spectrum, obtaining a master series of 5 spectra. The log of the observations is reported in Table~\ref{tab:log_duPont}.

\begin{table}
 \begin{center}
\caption{Log of du Pont observations, and S/N per pixel of the \ion{Ca}{ii} H\&K and \ion{Fe}{i} $\lambda$3928.0827 line cores.}\label{tab:log_duPont}
\scriptsize
  \begin{tabular}{cccccc}
 \hline\hline
Date & Orbital & Exp. time & S/N$_{\rm H}$ & S/N$_{\rm K}$ & S/N$_{\rm FeI}$\\
(MJD-55095) & phase & (s) & & & \\
\hline
14.506 & 0.368 & 840  & 197 & 169 & 158\\
15.506 & 0.691 & 5300 & 199 & 168 & 159\\
16.506 & 0.014 & 3800 & 134 & 98 & 86\\
17.506 & 0.338 & 5800 & 104 & 98 & 80\\
18.506 & 0.661 & 3800 & 114 & 90 & 78\\
\hline
  \end{tabular}
 \end{center}
\end{table}


\subsection{Analysis of the \ion{Ca}{ii} {\rm H\&K} lines} \label{sec:analysis}

In this section, we study the time variability of the chromospheric activity of \object{HD\,179949} through the \ion{Ca}{ii} H\&K diagnostics, searching for modulations phased with stellar rotation or the orbital motion of the planet. In the following, we define a proxy for the magnetic activity level and we describe the analysis of its time modulation.

\subsubsection{Extraction of the activity proxy}\label{sec:extraction}

To look for night-to-night variations in the core profiles due to chromospheric activity,  
first we compute both the H and K average spectral profiles for each series of observations (one per instrument), to be taken as representative of the \lq\lq average star\rq\rq. In the computation, we exclude the spectra between MJD=55111.5 and 55114 (corresponding to October 6 and 8 2009), because our simultaneous X-ray data (see Sect.~\ref{sec:xdata}) indicate enhanced flaring activity between these dates.

Then, we compute pixel-by-pixel differences of each normalized spectrum in the series from the corresponding high-S/N average spectrum (Fig.~\ref{fig:residues}). These differences are generally consistent with statistical noise, except for the core of the lines, where we find larger deviations with respect to the average spectrum. This is a consequence of the fact that the cores of the \ion{Ca}{ii} H\&K lines are sensitive to chromospheric activity. Next, we define the line core as the spectral range showing the reversal, and more precisely as a $\sim$1.5~\AA\ window centered on the line center, which includes the wavelength range over which we find significant deviations from the average spectrum.

In order to quantify a proxy for stellar activity, we integrate the difference spectra over the wavelength ranges defined above. The uncertainties on these \lq\lq Integrated Deviations\rq\rq\ (IDs hereafter) are computed by propagating the noise from the corresponding spectra. We remark that wider integration domains do not provide significantly different integrals, as the residuals outside the cores have null averages; the only effect would be the introduction of additional noise in the result.

We also note that the cross-correlation in wavelength of the spectra ensures IDs which are free from Doppler shifts due to the orbital motion of the star, and from possible inaccurate wavelength solutions due to instrumental misalignment of the spectra. In particular, if any misalignment among the spectra were present, we would find a systematic pattern in the residuals plotted in Fig.~\ref{fig:residues} at the positions of the narrow photospheric lines in the wings of the H\&K lines. Since the residuals away from line centers are consistent with randomly distributed noise fluctuations, this confirms the validity of the cross-correlation described in Sect.~\ref{sec:espadons}.

\subsubsection{Time series analysis}
\label{sec:time}

\begin{figure}
\centering
\includegraphics[width=\linewidth]{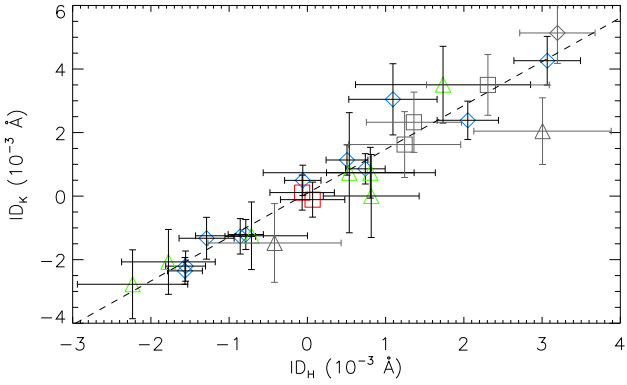}
\caption{Correlation between the ID$_{\rm H}$ and ID$_{\rm K}$ indicators. Blue diamonds, green triangles and red squares correspond to ESPaDOnS, SARG and du Pont measurements,  respectively. The dashed line is the outlier-resistant best fit of the trend. Symbols in gray correspond to spectra collected during three days characterized by an enhanced activity level, including a prominent X-ray flare (see Sect.~\ref{sec:xdata}), that were not included in the correlation analysis.}\label{fig:hk}
\end{figure}

The IDs described in the previous section are computed on a normalized flux scale. This approach allows us to merge the time series provided by the three sets of observations into one time series per line, consisting of 26 data points. Fig.~\ref{fig:hk} shows that the two proxies, ID$_{\rm H}$ and ID$_{\rm K}$, are well correlated (Pearson's coefficient $r\geq0.9$ and $p$-values close to 100\%). This result allows us to sum the two proxies, thus obtaining a merged time series of 26 data points with a higher S/N, which takes into account the contribution of both lines. Six of these data points were obtained during a period of enhanced flaring activity, as indicated by the X-ray data (Sect.~\ref{sec:xdata}). Since the search for a periodic signal can be affected by the presence of isolated flares, we first analyze the merged time series by excluding the corresponding data points. We also perform the analysis of the full series in Sect.~\ref{sec:flare}.

\begin{figure}
\centering
\includegraphics[width=\linewidth]{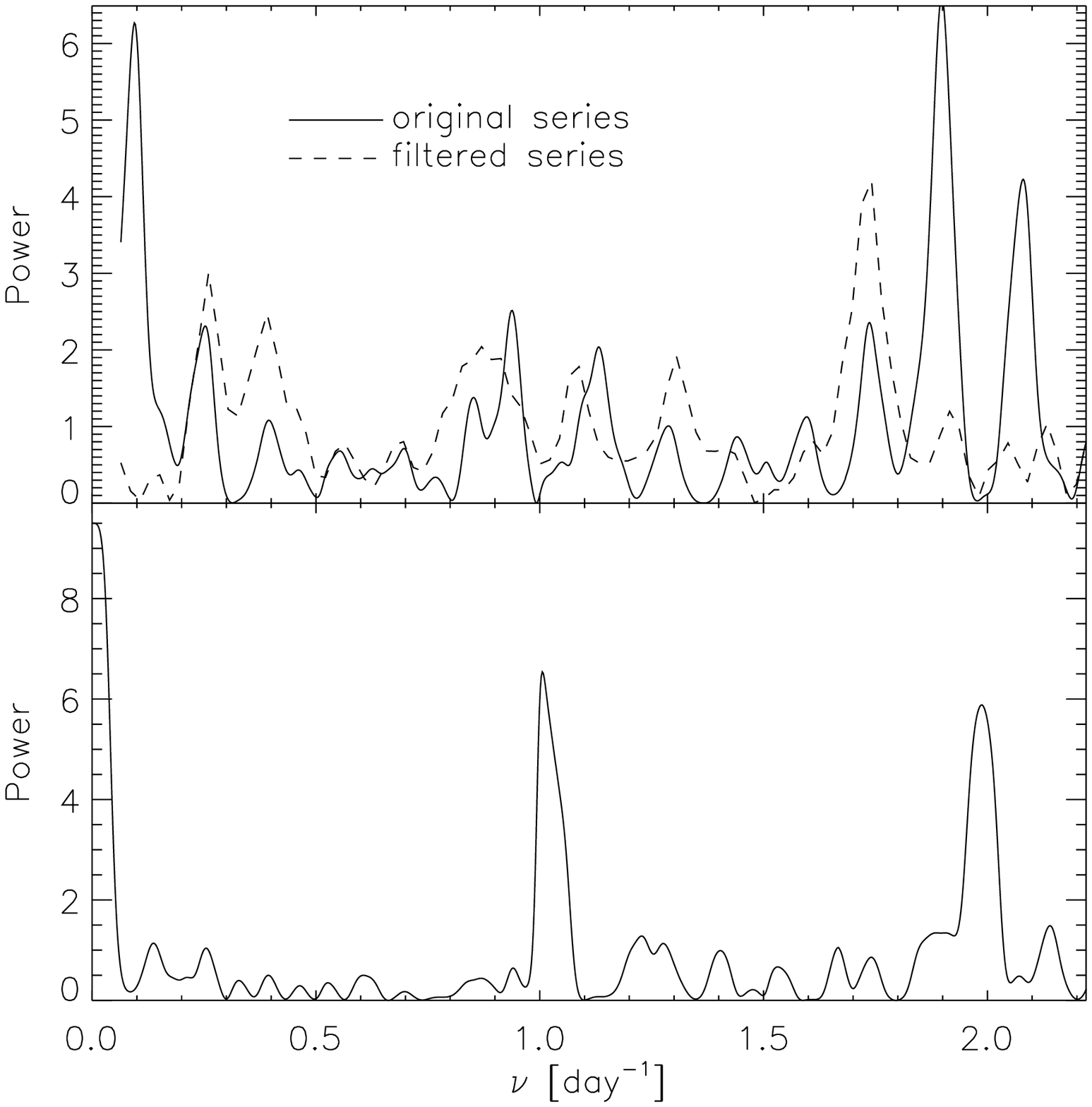}
\caption{\textit{Top panel - }Lomb-Scargle periodogram of the analyzed time series. The solid and dashed lines represent the periodogram for the original and frequency-filtered series, respectively. \textit{Bottom panel - } Spectral window of the analyzed time series.}\label{fig:scargle}
\end{figure}

As a preliminary step, we compute the Lomb-Scargle periodogram of the time series (Fig.~\ref{fig:scargle}, top panel). We find the highest power at a frequency $\nu_1 \sim 0.1$~day$^{-1}$, corresponding to a period of $\sim 10$~days, and a secondary peak at $\nu_2\sim 0.25$~day$^{-1}$ ($\sim 4$~days). To check whether the secondary frequency $\nu_{2}$ is an alias of $\nu_1$, we compute the spectral window of the series (Fig.~\ref{fig:scargle}, bottom panel). It looks quite noisy with significant sidelobes corresponding to the daily cadence of the observations and its harmonics, thus suggesting that aliasing cannot be responsible for the secondary peak at $0.25$~day$^{-1}$. To confirm this prediction, we fit our data with a trigonometric polynomial with frequency $\nu_1$ and subtracted it from the series.  Then we run the Lomb-Scargle analysis on the residuals, finding that the $\nu_2$ frequency is still present after filtering $\nu_1$. Thus, $\nu_2$ is not an alias of $\nu_1$.
 
\begin{figure*}
\centering
\includegraphics[width=.6\linewidth]{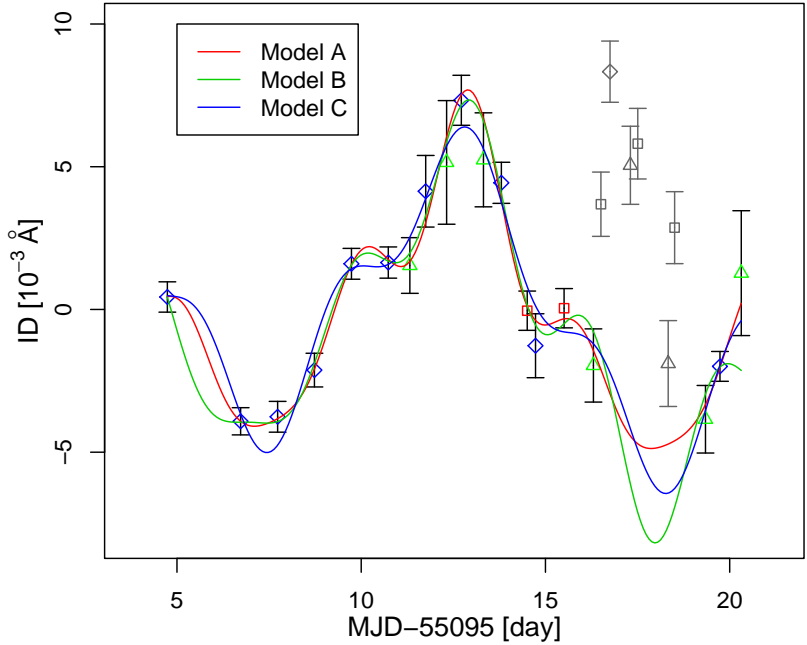}
\caption{Best-fits of the data adopting the models listed in Table~\ref{tab:model} and excluding the flaring activity interval. Solid red, green and blue lines represent the best fits of the models \textit{A}, \textit{B} and \textit{C} respectively. Symbols and colors of the data points are the same as in Fig.~\ref{fig:hk}.}\label{fig:model}
\end{figure*}

To obtain a more robust estimate of the periodicities in the time series, we fit the data with a model consisting of a sinusoidal expansion plus a linear trend, the latter representing any possible long-term variability:
\begin{equation}
ID=q+m\ t+ \sum_{n} A_n\sin\left(\omega_n\ t+\phi_n\right),
\label{eq:model}
\end{equation}
where $q$ and $m$ are the parameters of the linear trend, while $A_n$, $\omega_n$ and $\phi_n$ are the amplitude, angular frequency and initial phase of the $n$-th term of the sinusoidal expansion, respectively.

We use a non-linear least-squares estimation method to find the best-fit parameters \citep{Gallant1975}. The advantage of this approach over Scargle's analysis is that it takes into account the time variation of the data uncertainties, and it provides confidence intervals for the fitted parameters. Most importantly, it performs a fit of different periodic components by adjusting \textit{simultaneously} their parameters, while in Scargle's approach a multiperiodic variability can be detected and characterized only iteratively.

To analyze the time variations of the ID series, we make different assumptions on the origin of the variability, and we tune the model in Eq.~(\ref{eq:model}) accordingly. All the assumed models share the linear trend and the Fourier expansion of the stellar rotation, characterized by the pulsation $\omega_{\rm rot}= 2\pi/ P_{\rm rot}$, where $P_{\rm rot}$ is the rotation period. They differ in the inclusion of additional sinusoidal terms, chosen to match the various assumptions we wish to test. The models considered in our analysis are listed below and their properties are summarized in Table~\ref{tab:model}:
\begin{itemize}
	\item \textit{Model A}: it assumes that the time variations in the ID series are due to active regions in the stellar chromosphere, hard-rotating with an unknown period $P_{\rm rot}$; hence, in Eq.~(\ref{eq:model}) the $\omega_n$ parameters are the fundamental angular pulsation $\omega_{\rm rot}$ and its harmonics; 
		\item \textit{Model B}: it assumes that the signal is the sum of two contributions: the stellar rotation and a possible planet-triggered activity enhancement, phased with the orbital motion of the planet. In practice, we add to \textit{Model A} a secondary sinusoid with fixed period $P_{\rm sec}=P_{\rm orb}=3.092514$~days \citep[cf. ][]{2006ApJ...646..505B};
	\item \textit{Model C}: to investigate the possibility that the SPI signal is not exactly synchronized with the orbital motion of the planet, we add to \textit{Model A} a secondary sinusoidal term with a free period $P_{\rm sec}$.
\end{itemize}
To evaluate the optimal number of harmonics to be included in each model, 
 we adopt the corrected Akaike Information Criterion\footnote{The AIC is an extension of the canonical Likelihood-Ratio test for non-nested models.} (AICc), which provides a relative measure of the goodness of fit of different models when the number of data points is comparable with the number of free parameters \citep{Burnham2002}. This criterion not only rewards goodness of fit, but it also includes a penalty function that increases with the number of estimated parameters, thus discouraging overfitting.

In Table~\ref{tab:model} we summarize the output of the AICc: for each model we report the optimal number of harmonics $n_h$, the number of fitted parameters $n_p$\footnote{For \textit{Model A}, $n_p$ is given by the sum of two parameters from the linear trend ($q$ and $m$), the rotation period $P_{\rm rot}$, and two parameters for each harmonics (amplitude and initial phase), i.e., $n_p=2+1+2\cdot n$; for \textit{Model B} there are two additional parameters: the amplitude and the initial phase of the sinusoid with the period fixed at the orbital period of the planet; for \textit{Model C} there are three additional parameters: the amplitude, the period $P_{\rm sec}$, and the initial phase of the additional sinusoid.}, and the best fit periods. The AICc indicates that \textit{Model A} has the maximum likelihood, while \textit{Model B} and \textit{Model C} have confidence levels $<$1\% relative to \textit{Model A}. Moreover, a canonical $\chi^2$ test on \textit{Model A} gives a confidence level of $\sim$75\%. The best fits to the time series obtained with the different models are plotted in Fig.~\ref{fig:model}.
\begin{table}
 \begin{center}
\caption{Summary of  the best fit models.} 
\label{tab:model}
  \begin{tabular}{c|cccccc}
 \hline\hline
Model\tablefootmark{a} & $n_h$ & $n_p$ & $P_{\rm rot}$ & $P_{\rm sec}$ & AICc\tablefootmark{b} & $p_{\chi^2}$\tablefootmark{c}\\
 \hline
A & 4 & 11 & 10.8$\pm$0.2 & - & - & 76\%\\
B & 3 & 11 & 11.2$\pm$0.2 & - & $<$1\% & 43\%\\
C & 1 & 8 & 10.9$\pm$0.2 & 3.6$\pm$0.1 & $<$1\% & 13\%\\
\hline
A & 4 & 11 & 12.0$\pm$0.2 & - & 75\% & $<$1\%\\
B & 3 & 11 & 12$\pm$0.4 & - & $<$1\% & $<$1\%\\
C & 1 & 8 & 11$\pm$2 & 4.2$\pm$0.3 & - & $<$1\%\\
\hline
  \end{tabular}
 \end{center}
 \tablefoottext{a}{The first three lines refer to the censored time series (Sect.~\ref{sec:time}). The remaining lines refer to the full time series (Sect.~\ref{sec:flare}).}\\
\tablefoottext{b}{Relative confidence with respect to the selected model, as given by Akaike's criterion.}\\
\tablefoottext{c}{Classical $\chi^2$ likelihood.}
\end{table}

\begin{figure}
\centering
\includegraphics[viewport=30 15 425 255,clip,width=\linewidth]{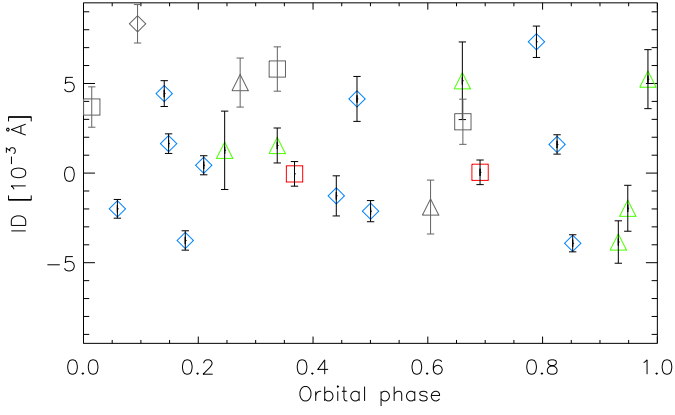}
\caption{ID series vs.\ the planetary phase. Symbols and colors are the same as in Fig.~\ref{fig:hk}.}\label{fig:ph_fold}
\end{figure}

This test significantly rejects models \textit{B} and \textit{C} with respect to \textit{Model A}. Specifically, we find that \textit{Model B}, i.e\ the model with $P_{\rm sec}$ locked at the planetary orbital period, can hardly be accepted as a good representation of the data. As a consequence, we exclude the scenario where the chromospheric activity is triggered or enhanced by the interaction with the planet and phased with its orbital motion. This is confirmed also by the phase folding of our activity proxy with the orbital period of the planet, as shown in Fig.~\ref{fig:ph_fold}: we do not find any systematic variation in our activity proxy with the phase of the orbital motion of the planet, contrary to, e.g., \citet{2003ApJ...597.1092S,2005ApJ...622.1075S} and \citet{Gurdemiretal12}.

\subsubsection{Analysis of the full time series}\label{sec:flare}
\begin{figure}
\centering
\includegraphics[width=\linewidth]{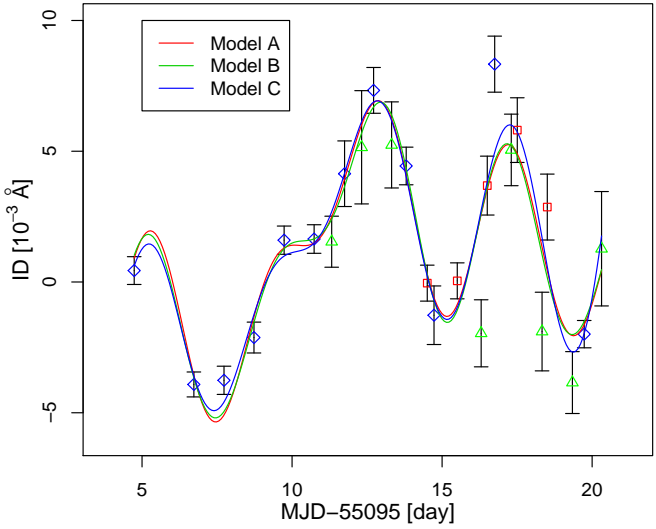}
\caption{Best-fits of the the full time series: the solid red, green and blue lines represent the best fits of models \textit{A}, \textit{B} and \textit{C} respectively. Symbols and colors of the data points are the same as in Fig.~\ref{fig:hk}.}\label{fig:model_flare}
\end{figure}

In the analysis described so far we have discarded the chromospheric flux data simultaneous with the X-ray flare that will be discussed in Sect.~\ref{sec:xdata}. We now relax this assumption and show the results of the analysis performed on the full time series. In principle, this is justified by the possibility that the flare be associated with the magnetic SPI \citep[cf., e.g.,][ and references therein]{Lanza12}. The best-fits with the same models considered in Sect.~\ref{sec:time} are shown in Fig.~\ref{fig:model_flare}. The goodness of the fit is poorer than for the censored time series, especially during the flaring time interval, and all fits fail to pass the classical $\chi^{2}$ significance test.  This is probably a consequence of the enhanced short-term variability associated with the flare rather than the effect of increasing the number of data points and corresponding decrease of the number of degrees of freedom. 

Nevertheless, if we assume that at least one of the models may still provide an acceptable fit to the data, we can apply the AICc approach to find the one with the best likelihood. With this presumption, the best model turns out to be \textit{Model C} including the principal and first harmonics of the rotational modulation. The two periods returned by the fit are $P_{\rm rot}$=11$\pm$2~days and $P_{\rm sec}$=4.2$\pm$0.3~days.

The second most likely model is \textit{Model A} expanded up to the second harmonics of $P_{\rm rot}$, whose best-fit estimate is 12.0$\pm$0.2~days. The relative confidence of \textit{Model A} with respect to \textit{Model C} is $\sim$75\%, i.e.\ we cannot reject \textit{Model A} if we accept \textit{Model C}. On the other hand, we can reject \textit{Model B}, as its confidence level relative to \textit{Model A} is close to zero, i.e.\ we can safely exclude the scenario of a SPI phased with the orbital motion of the planet. See Table~\ref{tab:model} for a summary of the fitted parameters.

We remark that models \textit{A} and \textit{C} provided similar results. As a matter of fact, the best-fit \textit{Model C} turns out with $P_{\rm sec}$ consistent with the second harmonic of $P_{\rm rot}$. Moreover, the estimates of $P_{\rm rot}$ returned by the two models are consistent within the uncertainties. In any case, we stress that both models provide poor fits to the full time series. This is also indicated by the fact that the uncertainties in the fitted parameters are larger than in the case of the censored time series considered in Sect.~\ref{sec:time}, i.e., with the time interval of the X-ray flare removed.

\subsubsection{Significance of the results}
\label{sec:significance}

\begin{figure}
\centering
\includegraphics[viewport=580 360 1110 680,clip,width=\linewidth]{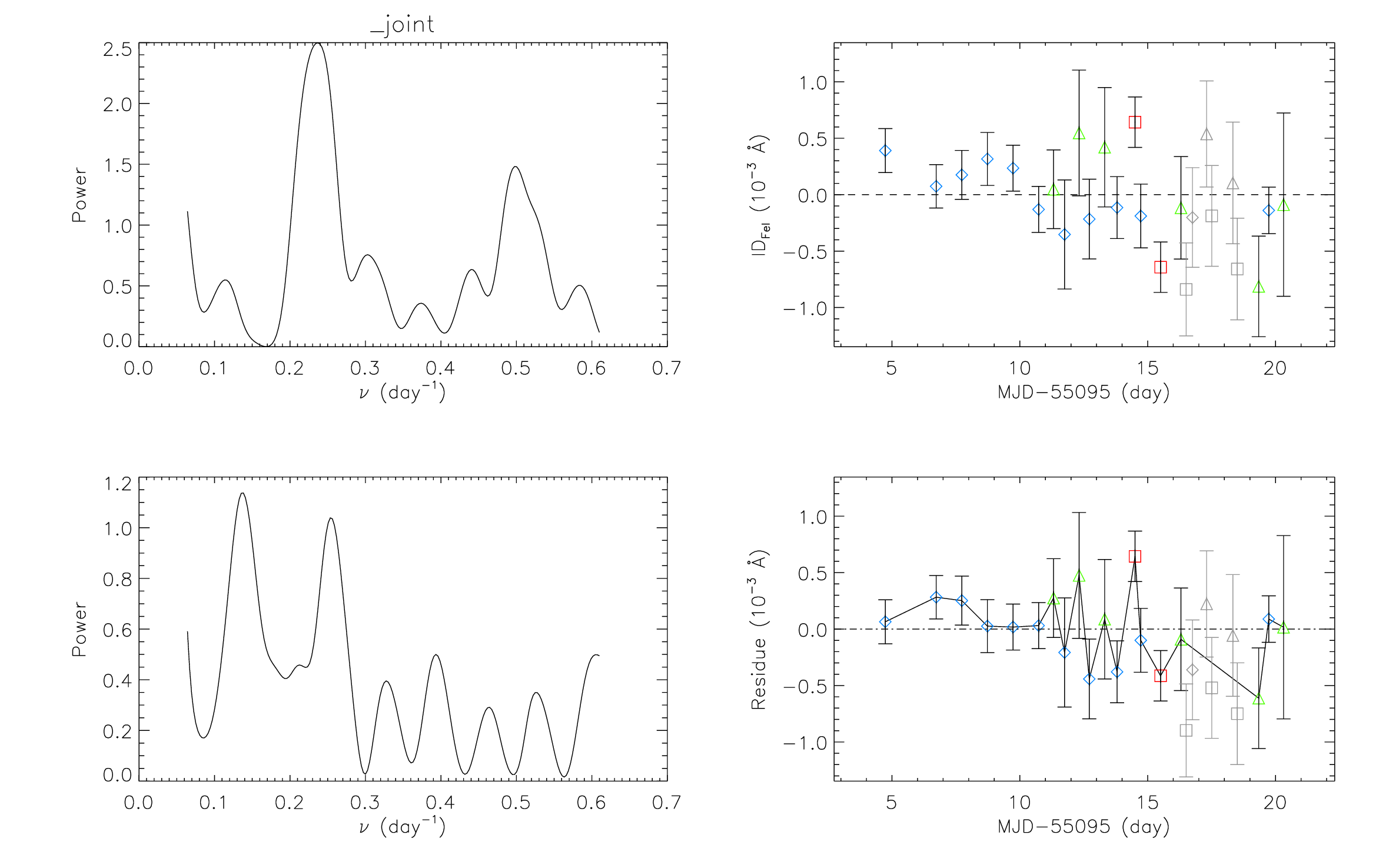}
\caption{ID time series for the \ion{Fe}{i} $\lambda$3928.0827 line: the derived deviations are consistent with the horizontal line at ID=0 and do not show any significant time variability. Symbols and colors are the same as in Fig.~\ref{fig:hk}.}\label{fig:FeI}
\end{figure}

To assess the robustness of our results, first we check whether the derived periodicities are due to instrumental effects. As a sanity check, we perform the same analysis on the \ion{Fe}{i} $\lambda$3928.0827 line, identified in the same echelle order as the \ion{Ca}{ii} K line. This line, being a photospheric feature, is supposed to be weakly affected by chromospheric activity. Therefore, we do not expect any modulation with the rotation of the star or with the orbital motion of the planet; if present, any modulation of its IDs is to be ascribed to statistical noise or, in the worst case, to instrumental or reduction effects.  By integrating the difference spectra over a 0.3~\AA\ width centered on the \ion{Fe}{i} line, we obtain the merged ID series shown in Fig.~\ref{fig:FeI}. The data points are consistent with a constant line at ID(t)=0  within the uncertainties, i.e. there is no evidence of any coherent modulation vs.\ time. We can thus safely exclude that the periodicities found in the \ion{Ca}{ii} H\&K lines stem from instrumental effects or flaws in the reduction procedures.

To estimate the significance of our results, we also run a simple test on our time series: we shuffle the flux and time measurements to obtain 10,000 fake data samples; then, for each realization, we compute the maximum-likelihood fit for \textit{Model A} and compute the periodicity $P_{\rm rot}$. This test aims at destroying the true periodicities of the series by time-shuffling and, by consequence, it returns only frequencies artificially introduced by either random noise or the actual time sampling of the observations.

\begin{figure}
\centering
\includegraphics[width=\linewidth]{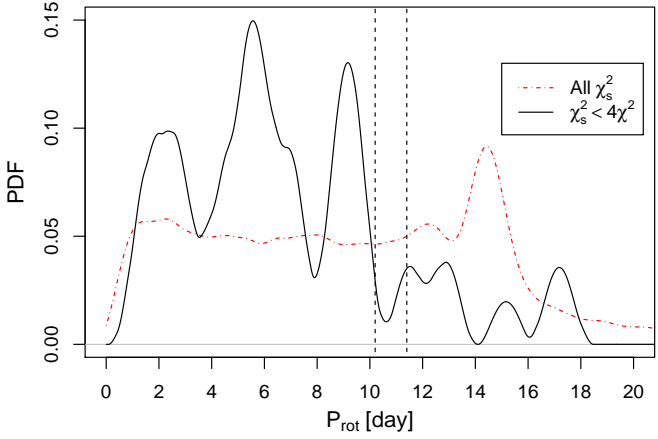}
\caption{Probability Distribution Function of the period $P_{\rm rot}$ as recovered from 10,000 realizations by time-shuffling, considering all the cases (solid black line) or the realizations returning a $\chi^{2}_{\rm s}$ lower than 4 times the best-fit $\chi^{2}$ for \textit{Model A} (dot-dashed red line; see Sect.~\ref{sec:significance}). The dashed vertical lines bracket the 3$\sigma$ interval around the best-fit period of $10.8$~days.}\label{fig:fap}
\end{figure}

In Fig.~\ref{fig:fap}, we plot the derived Probability Distribution Function (PDF) of the parameter $P_{\rm rot}$, as derived by fitting \textit{Model A} and considering two cases: the dot-dashed red PDF has no constraint on the final value of the $\chi^{2}_{\rm s}$ of the best fit, thus including also models having a goodness of fit much worse than we obtained with \textit{Model A}; the solid black-lined PDF includes only those models that have a $\chi_{\rm s}^{2}$ lower than four times the $\chi^{2}$ of \textit{Model A}: this threshold aims at reducing the impact of the low-quality best fits on the PDF. 

When we apply the constraint on the $\chi^{2}$ value, the PDF shows a significant value for periods longer than $\sim 1.0$~days, with peaks corresponding to an integer or semi-integer number of days. The period of $1.0$~day corresponds to the first sidelobe of the spectral window (cf. Fig.~\ref{fig:scargle}, lower panel), while the other periodicities correspond to an integer or semi-integer number of days, i.e., the subharmonics of the first sidelobe, as a consequence of the daily cadence of the subsets of observations obtained from the same observatory. 

To get an estimate of the False Alarm Probability (FAP), we integrate the constrained PDF over the 3$\sigma$ interval centered on each best fit value, which corresponds to the $p\gtrsim$99\% confidence interval. The probability that random noise and/or time sampling lead to a period consistent within 3$\sigma$ with the best-fit estimate of $P_{\rm rot}$ turns out to be FAP$\sim$2.2\%. Similarly, the probability that a periodicity in the range $3.5-4$~days randomly arises from the noise present in the time series is FAP$\lesssim$3\%.

\begin{figure*}
\centering
\includegraphics[viewport=1 1 545 815,clip,width=.3\linewidth]{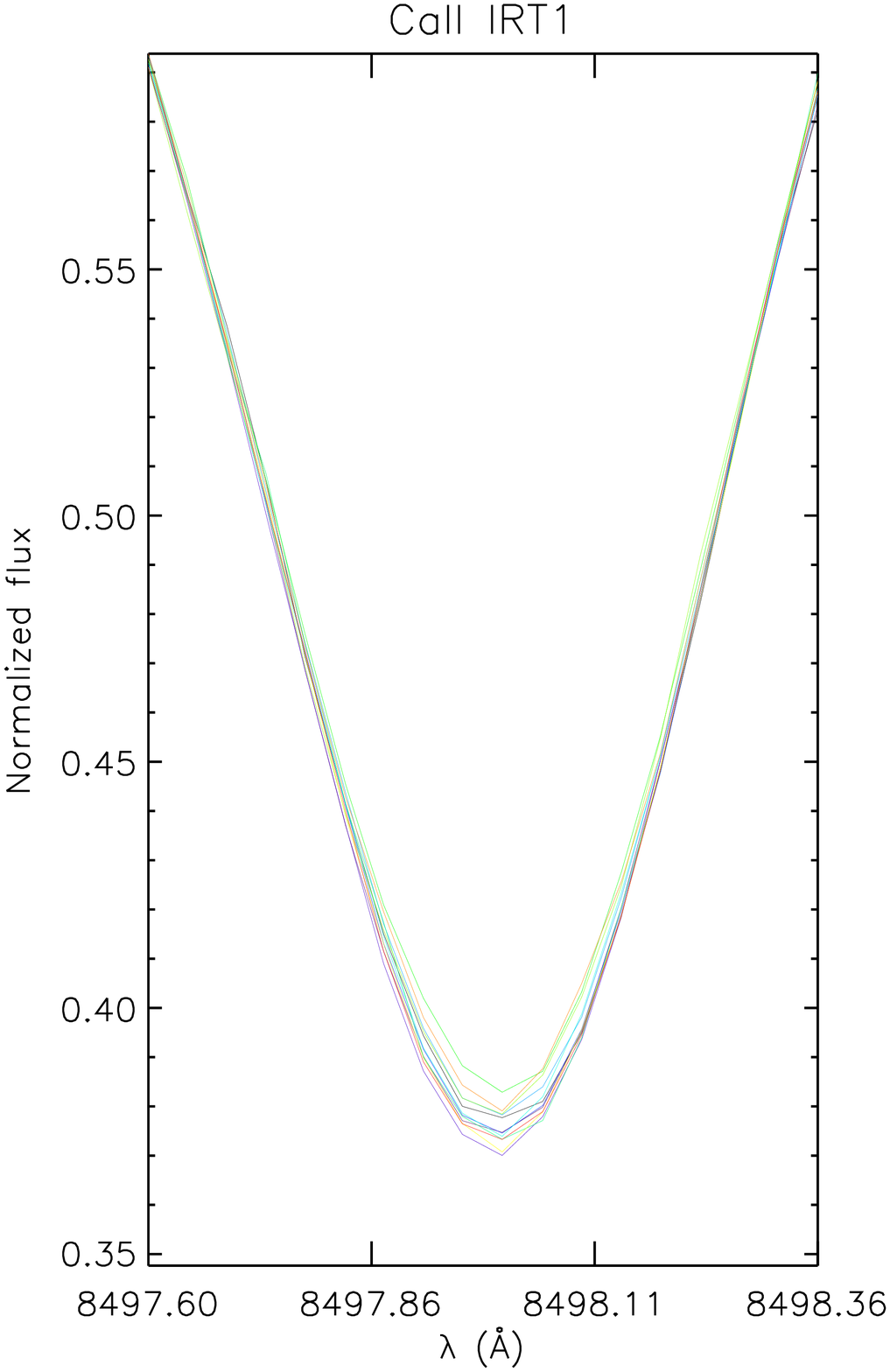}
\includegraphics[viewport=1 1 545 815,clip,width=.3\linewidth]{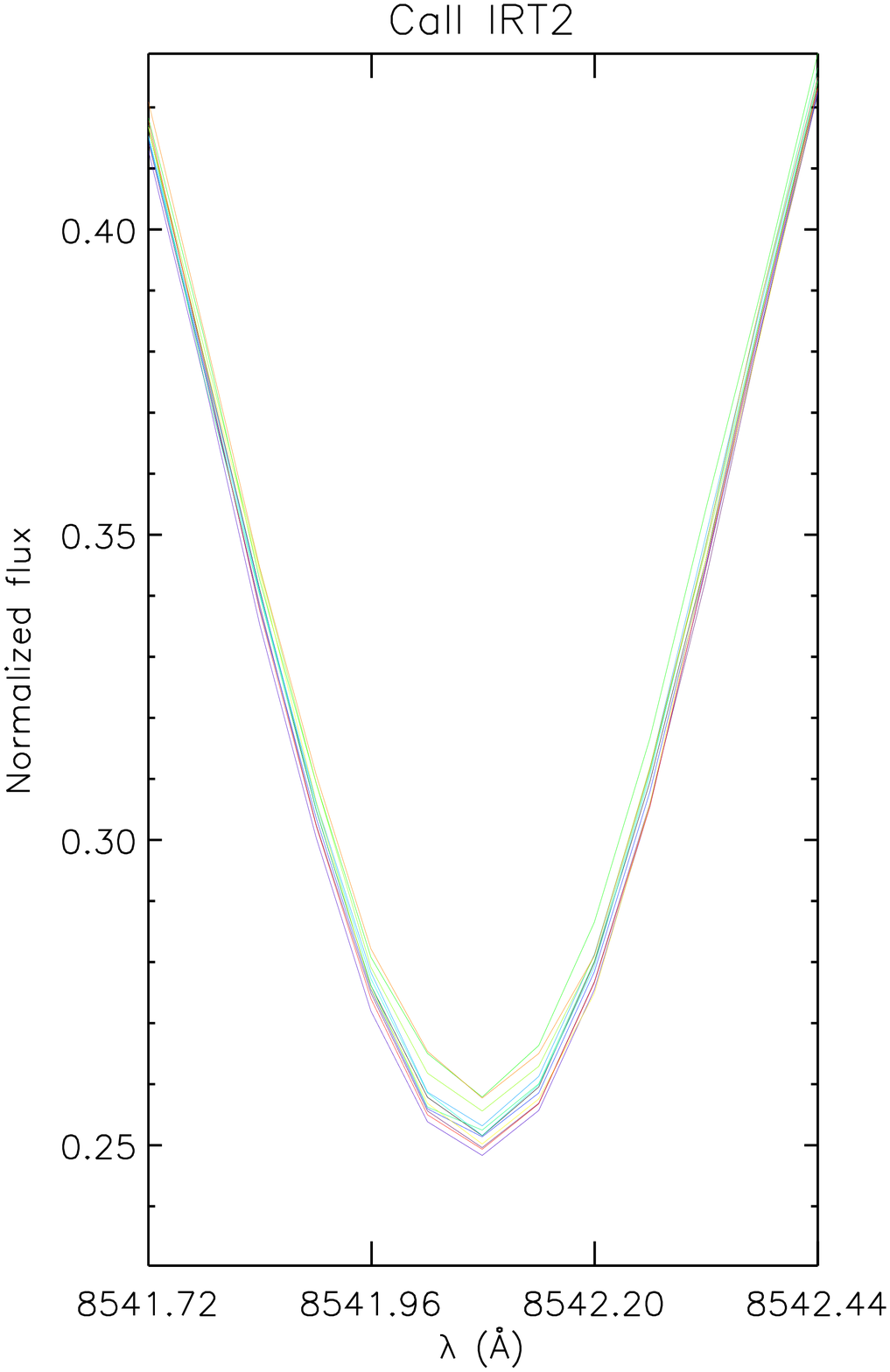}
\includegraphics[viewport=1 1 545 815,clip,width=.3\linewidth]{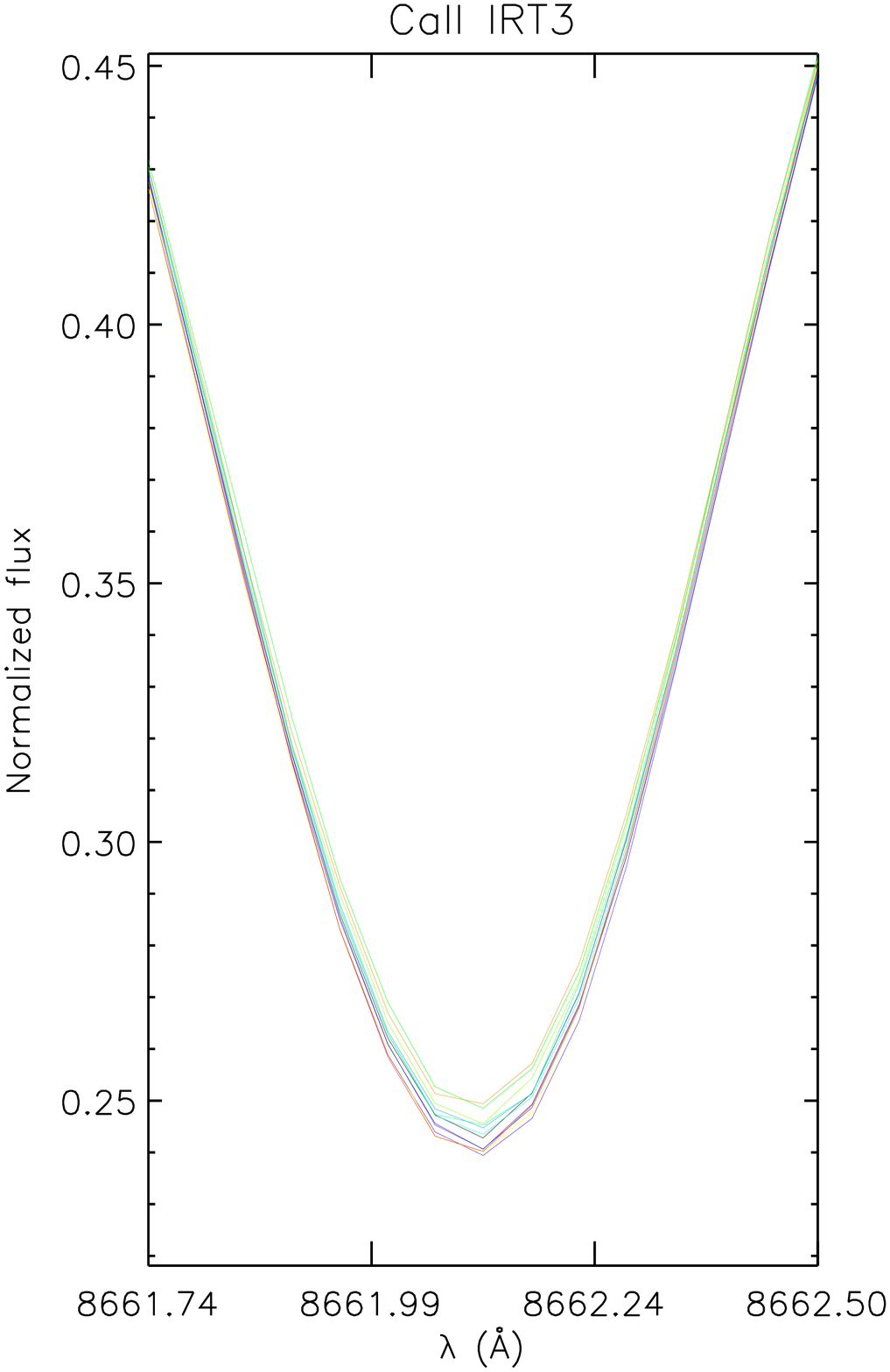}
\caption{Spectra of the line cores of the \ion{Ca}{ii} IRT of \object{HD\,179949} taken with ESPaDOns. Different colors code the nights of observation.}\label{fig:irt}
\end{figure*}

\subsection{Analysis of the \ion{Ca}{ii} {\rm IRT}}\label{sec:irt}

The ESPaDOnS spectra also cover the spectral range containing the \ion{Ca}{ii} Infra-Red Triplet (IRT) located at $\lambda$8498, 8542, 8662~\AA. One could expect that the low contrast between the variations in the line cores due to chromospheric activity and the continuum prevents the analysis of any temporal variability. Nonetheless, in this spectral region the number of photon counts is remarkably higher than the counts in the H\&K spectral region (i.e.\ the blue side of the spectrum). Indeed, the S/N in the core of the IRT lines is higher (typically S/N$\geq$250) than the S/N of the H\&K lines (see Tables~\ref{tab:log_espadons}, \ref{tab:log_sarg}, and~\ref{tab:log_duPont}).

Moreover, variations in the line profiles of the IRT have already been studied by \citet{2008ApJ...676..628S}, finding that the IRT-based activity proxies are correlated with the H\&K proxies. We therefore analyzed the lines in the IRT, and their temporal evolution, in the same way as we analyzed the \ion{Ca}{ii} H\&K lines in Sect.~\ref{sec:data} and \ref{sec:analysis}.

\begin{figure}
\centering
\includegraphics[viewport=1 480 545 790,clip,width=\linewidth]{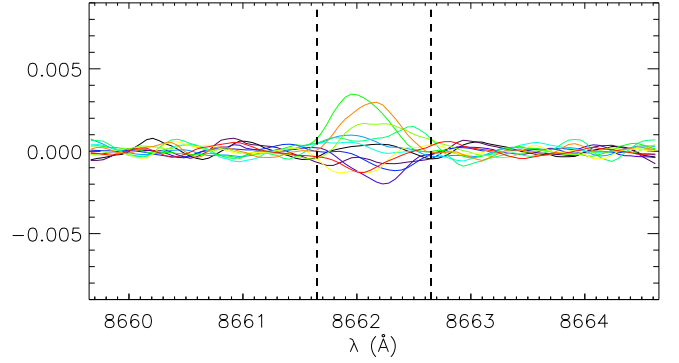}
\caption{Deviations from the average spectrum of the $\lambda$8662~\AA\ line. Colors code the different nights of observations. The vertical dashed lines delimit the core of the lines, as assumed in the present analysis.}\label{fig:irt_residue}
\end{figure}

\begin{figure}
\centering
\includegraphics[width=\linewidth]{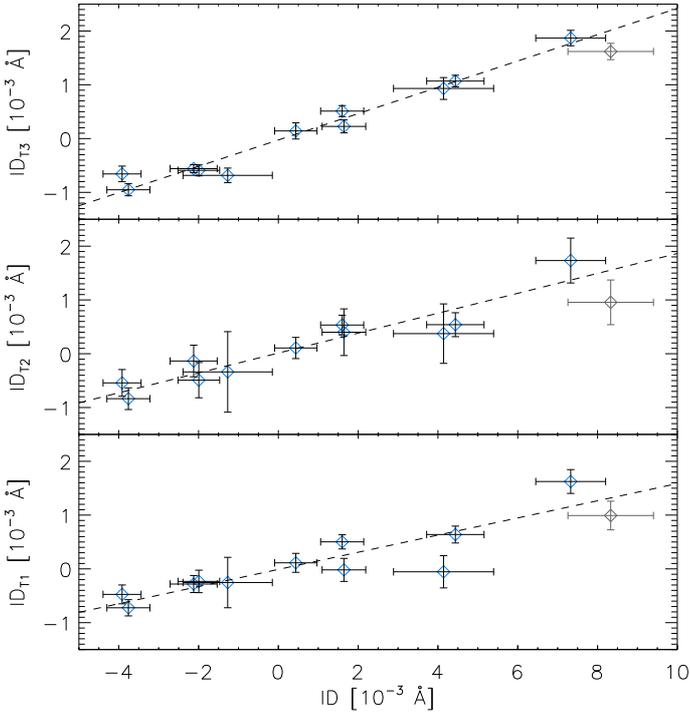}
\caption{Correlation between the co-added ID of the H\&K lines and the IDs of the three lines in the \ion{Ca}{ii} IRT (T1, T2, and T3, from bottom to top). In each panel, the dashed line is the outlier-resistant best fit of the trend (Pearson's coefficient always $\gtrsim$0.9). Symbols in gray correspond to the spectra collected during the flaring interval indicated by the X-ray observations (Sect.~\ref{sec:xdata}), and not included in the correlation analysis.}\label{fig:irt_corr}
\end{figure}

In Fig.~\ref{fig:irt} we show the time series of the spectra of the three components of the triplet (named T1, T2 and T3, from the bluest to the reddest component) as taken by ESPaDOnS, while in Fig.~\ref{fig:irt_residue} we plot, as an example, the residuals with respect to the average spectrum for the $\lambda$8662~\AA\ line. We find that the residuals integrated over $\sim$0.5~\AA\ around the line center are not consistent with random noise, and are correlated with the ID of the H\&K lines defined in Sect.~\ref{sec:time} (Fig.~\ref{fig:irt_corr}).

\begin{figure}
\centering
\includegraphics[width=\linewidth]{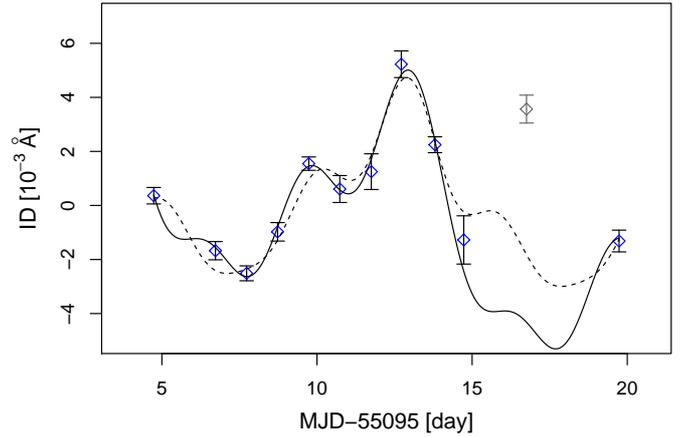}
\caption{Best-fit (solid line) for the ID proxy of the \ion{Ca}{ii} IRT with Eq.~(\ref{eq:model}) \textit{Model A}, compared with the best-fit of the H\&K data (dashed line, see text for more details). The gray symbol represents the data point not included in the fitting procedure, being likely contaminated by the flare activity as discussed in Sect.~\ref{sec:xdata}.}\label{fig:irt_fit}
\end{figure}

We perform the same analysis as described in Sect.~\ref{sec:time}. We summed the ID series of the three lines, and fit the co-added series with \textit{Model A} (Table~\ref{tab:model}) using the non linear least-squares algorithm (Fig.~\ref{fig:irt_fit}): the best-fit $P_{\rm rot}$ is fully consistent within the error bars with the one obtained in Sect.~\ref{sec:time}.

For the sake of comparison, we overplot in Fig.~\ref{fig:irt_fit} the best fit to the H\&K data, fixing the frequencies and scaling the amplitudes of the fitted sinusoids so to minimize the sum of the squared residuals in the case of the \ion{Ca}{ii}~IRT indicator. The main difference between the fitted model and the scaled model is that the former more closely follows the data points, while the latter gives a smoother fit. This is due to the fact that the fit is based on a lower number of data points. As a consequence, the fitted model adjusts itself to each single data point rather than to the general trend, as it is the case of the scaled model. Nonetheless, the fitted model still reproduces the general trend of the scaled model, and this is a clear indication that chromospheric variability could be alternatively performed on IRT line measurements if provided by ad hoc observational campaigns.

\section{X-ray data}\label{sec:xdata}

\object{HD\,179949} was observed 15 times with XMM-Newton, between Sep 21 and Oct 27, 2009 (see the observation log in Table~\ref{tab:xlog}). The nominal exposure times ranges from about 16\,ks to 68\,ks, depending on the actual short-term XMM-Newton schedule in the AO8 observing season. On Sep 21, Oct 3, and Oct 11, 2009, strictly simultaneous observations with SARG were performed, while in three other occasions (Oct 1, 2, and 6) optical spectroscopy with the same instrument was performed within a few hours from the X-ray observations. Similarly, simultaneity was achieved on Oct 2 and Oct 6 with CFHT, and on Oct 4, 5, and 6 with du Pont. We employed the {\sc mos} and {\sc pn} detectors in {\sc full window} mode and with the {\sc medium} filter in all observations.

We reprocess all the data with the standard pipeline (SAS V9.0), and perform a time screening for high background intervals (proton flares) that are removed from the following analysis. Only four observations are actually affected by such a problem (Table~\ref{tab:xlog}). We then extract {\sc mos} and {\sc pn} source counts from circles of 34.5\arcmin\ and 36\arcmin\ radius, respectively, while the local background is sampled from a nearby non-overlapping circle of 85\arcmin\ radius in all cases. Background-subtracted count rates are reported in Table~\ref{tab:xlog} for each observation and instrument.

\begin{table*}
\caption{Log of the {\it XMM-Newton} observations of \object{HD\,179949}.}
\label{tab:xlog}
\footnotesize
\begin{center}
\begin{tabular}{cccc@{\hspace{1mm}}ccr@{\hspace{1mm}}ccl}
\hline\hline
&  Start date & \multicolumn{1}{c}{Time\tablefootmark{a}}  & \multicolumn{3}{c}{MOS\tablefootmark{b}}  & \multicolumn{3}{c}{pn}  \\
& (2009) & MJD & \multicolumn{1}{c}{Exp\tablefootmark{c}} & & Rate & Exp\tablefootmark{c} & & Rate  \\
ObsId & (UT) & (d) & \multicolumn{1}{c}{(ks)} & & (cnt/ks) & \multicolumn{1}{c}{(ks)}
& & (cnt/ks) & Note\\
\hline
Obs1  & Sep 21 21:08 & 55095.99 & 18.25 &   & $36.9 \pm 1.5$ & 14.62 &   & $183 \pm 4$ 
& Simultaneous TNG obs\\
Obs2  & Sep 23 21:17 & 55098.07 & 30.76 &   & $51.4 \pm 1.3$ & 25.53 &   & $251 \pm 3$ \\
Obs3  & Oct 01 14:31 & 55105.70 & 16.38 &   & $48.7 \pm 1.8$ & 12.97 &   & $228 \pm 4$ 
& TNG obs 40\,m later \\
Obs4  & Oct 02 01:57 & 55106.18 & 16.37 &   & $54.4 \pm 1.9$ &  7.40 & c & $254 \pm 6$ 
& simultaneous CFHT obs\\
Obs5  & Oct 02 23:46 & 55107.07 & 13.61 &   & $41.5 \pm 1.8$ & 10.56 &   & $201 \pm 4$ 
& TNG obs 4\,h earlier; CFHT obs 2\,h later \\
Obs6  & Oct 03 15:40 & 55107.95 & 49.85 &   & $43.4 \pm 1.0$ & 42.29 &   & $202 \pm 2$ 
& Simultaneous TNG obs; CFHT obs 2\,h later\\
Obs7  & Oct 04 23:37 & 55109.12 & 21.85 & c & $38.8 \pm 1.4$ & 17.74 & c & $189 \pm 3$ 
& Simultaneous du Pont obs; CFHT obs 1\,h later\\
Obs8  & Oct 05 14:17 & 55109.99 & 67.10 &   & $38.8 \pm 0.8$ & 47.45 & c & $189 \pm 2$ 
& Simultaneous du Pont obs\\
Obs9  & Oct 06 23:20 & 55111.22 & 37.14 & c & $52.7 \pm 1.2$ & 30.86 & c & $255 \pm 3$ 
& TNG obs 4\,h earlier; simultaneous du Pont and CFHT obs \\
Obs10 & Oct 11 15:51 & 55115.78 & 18.94 &   & $47.6 \pm 1.6$ & 15.22 &   & $219 \pm 4$
& Simultaneous TNG obs, but low-quality data\\
Obs11 & Oct 14 22:25 & 55119.09 & 25.24 &   & $45.8 \pm 1.4$ & 20.72 &   & $214 \pm 3$ \\ 
Obs12 & Oct 17 07:08 & 55121.40 & 17.36 &   & $39.5 \pm 1.5$ & 13.83 &   & $198 \pm 4$ \\
Obs13 & Oct 19 09:04 & 55123.49 & 19.42 &   & $42.6 \pm 1.5$ & 15.64 &   & $212 \pm 4$ \\
Obs14 & Oct 22 22:21 & 55127.07 & 23.26 &   & $48.5 \pm 1.5$ & 19.01 &   & $237 \pm 4$ \\
Obs15 & Oct 27 19:13 & 55131.90 & 15.39 &   & $47.6 \pm 1.8$ & 12.12 &   & $226 \pm 4$ \\
\hline
\end{tabular}
\tablefoot{
\tablefoottext{a}{Central time of the observing window.}
\tablefoottext{b}{MOS2 data; MOS1 data are almost identical.}
\tablefoottext{c}{Net exposure time after screening for high background levels.}
}
\end{center}
\end{table*}
\normalsize

\begin{figure*}
\centering
\includegraphics[viewport=1 1 545 495,clip,width=.8\textwidth]{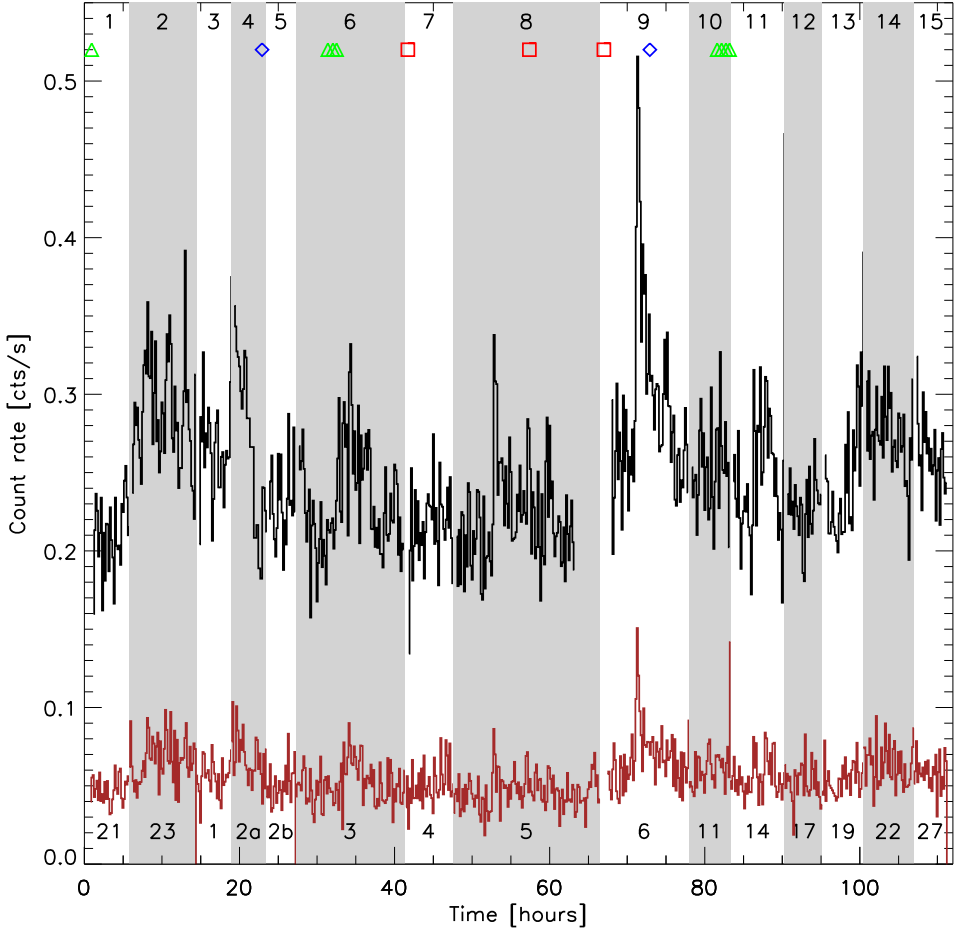}
\caption{X-ray light curves for the 15 XMM-Newton observations (numbered on top), artificially joined in time, with dates in September and October 2009 indicated at the bottom. The cumulative exposure time since the start of the first observation is reported in abscissa. The upper (black) curve shows the {\sc pn} count rate, while the lower (red) curve refers to the {\sc mos1} detector. All time bins have the same nominal duration of 600\,s. The symbols on top indicate the strictly simultaneous optical spectroscopy  (triangles for SARG, squares for du Pont, diamonds for ESPaDOnS observations).  
}
\label{fig:lc}
\end{figure*}

We start the X-ray data analysis extracting X-ray light curves with a constant bin size of 600\,s: in Fig.~\ref{fig:lc} we artificially join in time all these light curves to give an overall idea of the source variability during one month of observing time. The source showed in fact variability by less than a factor 4 overall, on short (hours) and long (days-weeks) time scales, including a prominent flare detected on October 6 2009 (Obs 9). This variability appears to be typical of late-type stars with an intermediate coronal activity level, and it is difficult to identify isolated flaring events other than in the case of Obs 9. In the same figure we marked the times of the strictly simultaneous SARG, ESPaDOnS and du Pont observations.

Then we perform a spectral analysis of each observation, in order to evaluate the coronal plasma temperature and the X-ray luminosity. The total number of counts in each {\sc pn} spectrum ranges from $\approx$ 2700 to 7900, and in the range 700--2000 counts for each of the {\sc mos} spectra. With the available photon counting statistics, two-component isothermal (2-T) models are adequate to describe the source X-ray emission, and to constrain the interstellar absorption. In practice, we employ the spectra with the highest number of total counts (Obs 6, 8, and 9) to check the hydrogen column density toward the source, $N_{\rm H}$, and the abundances of some individual elements in the emitting plasma. The $N_{\rm H}$ turns out to be low and compatible with the value expected for typical values of the interstellar density \citep{Savage1979} and a source at the distance of \object{HD\,179949} ($d \simeq 27$~pc). In all the subsequent spectral fitting, we fixed $N_{\rm H} = 10^{19}$\,cm$^{-2}$. 

Regarding the chemical composition of the coronal plasma, we check that non-solar abundances are required in order to obtain good quality fits of the observed spectra. In particular, the iron abundance is about half solar \citep[using][as reference]{Anders1989}, while the stellar photosphere is metal-rich (\citet{Valenti2005} [Fe/H]=0.14; \citet{Sousa2008} Fe/H=[0.21]). The spectra with the best signal-to-noise ratio also show that non-solar abundance ratios are required to fit properly some prominent line emission complexes: in fact, the best fitting models for Obs 6, 8, and 9 suggest O/Fe = 0.3, Ne/Fe = 0.5, and Si/Fe = 0.9, while all the other elements have abundances essentially equal to the iron one. The iron underabundance by a factor of 3 in corona with respect to the photospheric composition is similar to the well-studied case of $\tau$~Bootis, another planet-hosting star \citep{Maggio2011}.

After this preliminary analysis, we decide to fix the abundance ratios to the value reported above, and to perform 2-T spectral fitting with the iron abundance also fixed to 0.5 solar\footnote{We verified that the best-fit Fe abundance, if left free to vary, is comprised in the narrow range 0.4--0.6 in solar units.}. Finally, given the different exposure times of the XMM-Newton observations, we split the 4 longest observations (Obs2, Obs6, Obs8, and Obs9) in two or three segments so to increase the time resolution of the spectral analysis and to ensure a number of total counts similar to those in the other observations. This choice also ensures similar statistical uncertainties in the best-fit parameters. The final results are reported in Table~\ref{tab:fit}.

\begin{table*}
\caption{X-ray spectral fitting results\tablefootmark{a}.}
\label{tab:fit}
\footnotesize
\begin{center}
\begin{tabular}{ccccclccc}
\hline\hline
& Orbital & $T_1$ & $\log EM_1$ & \multicolumn{1}{c}{$T_2$} & & $L_{\rm x}$\tablefootmark{b} \\
ObsId & phase & (MK) & ($cm^{-3}$) & \multicolumn{1}{c}{(MK)} & \multicolumn{1}{c}{$EM_1/EM_2$} &
($10^{28}$\,erg/s) & \multicolumn{1}{c}{$\chi^2_{red}$} & dof \\
\hline
Obs1  & 0.160 & $3.3^{+0.3}_{-0.4}$ & 51.4 & $6.2^{+0.5}_{-1.8}$ & 3.4 &
$2.9^{+1.1}_{-0.2}$ & 1.4 & 55 \\
Obs2a  & 0.804 & $3.8^{+0.1}_{-0.1}$ & 51.6 & $8.7^{+0.9}_{-1.5}$ & 6.7 &
$3.9^{+0.2}_{-0.2}$ & 1.3 & 65 \\
Obs2b  & 0.862 & $2.9^{+0.3}_{-0.4}$ & 51.3 & $6.5^{+0.3}_{-0.4}$ & 1.2 &
$3.9^{+0.2}_{-0.2}$ & 1.3 & 66 \\
Obs3  & 0.300 & $3.8^{+0.1}_{-0.2}$ & 51.5 & $7.9^{+1.1}_{-1.3}$ & 10.4 &
$3.5^{+0.2}_{-0.2}$ & 1.6 & 57 \\
Obs4  & 0.454 & $3.1^{+0.4}_{-0.5}$ & 51.4 & $6.3^{+0.4}_{-0.6}$ & 1.5 &
$4.0^{+0.2}_{-0.2}$ & 1.5 & 48 \\
Obs5  & 0.743 & $3.4^{+0.6}_{-0.5}$ & 51.4 & $5.9^{+1.8}_{-1.8}$ & 3.3 &
$3.1^{+1.9}_{-0.1}$ & 1.4 & 42 \\
Obs6a  & 0.966 & $3.4^{+0.5}_{-0.4}$ & 51.4 & $5.8^{+1.0}_{-1.0}$ & 2.6 &
$3.0^{+1.3}_{-0.1}$ & 1.1 & 58 \\
Obs6b  & 0.028 & $3.4^{+0.4}_{-0.5}$ & 51.4 & $6.8^{+0.9}_{-0.5}$ & 2.4 &
$3.6^{+0.3}_{-0.2}$ & 0.9 & 57 \\
Obs6c  & 0.088 & $3.6^{+0.3}_{-0.8}$ & 51.4 & $6.3^{+1.4}_{-1.1}$ & 3.8 &
$3.0^{+0.2}_{-0.2}$ & 1.3 & 58 \\
Obs7  & 0.407 & $3.3^{+0.3}_{-0.3}$ & 51.4 & $6.3^{+0.6}_{-0.8}$ & 3.1 &
$3.0^{+0.3}_{-0.2}$ & 1.2 & 69 \\
Obs8a  & 0.599 & $2.8^{+0.3}_{-0.5}$ & 51.2 & $5.5^{+0.6}_{-0.3}$ & 1.2 &
$3.0^{+0.2}_{-0.2}$ & 1.1 & 62 \\
Obs8b  & 0.674 & $2.3^{+0.7}_{-0.5}$ & 51.1 & $5.6^{+0.6}_{-0.3}$ & 0.7 &
$3.2^{+0.2}_{-0.2}$ & 1.2 & 64 \\
Obs8c  & 0.762 & $3.7^{+0.3}_{-2.0}$ & 51.5 & $6.2^{+1.5}_{-4.4}$ & 8.4 &
$2.9^{+0.5}_{-0.2}$ & 1.6 & 61 \\
Obs9a  & 0.047 &  $3.6^{+0.3}_{-0.6}$ & 51.5 & $7.1^{+1.2}_{-1.1}$ & 4.1 &
$3.7^{+0.3}_{-0.2}$ & 1.2 & 53 \\
Obs9b  & 0.093 & $3.4^{+0.4}_{-0.9}$ & 51.4 & $7.0^{+1.8}_{-0.5}$ & 1.5 &
$4.5^{+0.4}_{-0.3}$ & 0.9 & 53 \\
Obs9c  & 0.137 & $3.7^{+0.2}_{-0.4}$ & 51.5 & $7.4^{+0.9}_{-0.8}$ & 3.8 &
$3.7^{+0.3}_{-0.2}$ & 1.5 & 53 \\
Obs10 & 0.566 & $3.0^{+0.6}_{-0.4}$ & 51.4 & $5.8^{+1.6}_{-0.4}$ & 1.5 &
$3.5^{+0.2}_{-0.2}$ & 1.4 & 70 \\
Obs11 & 0.627 & $2.9^{+0.9}_{-0.6}$ & 51.3 & $6.0^{+0.6}_{-0.5}$ & 1.4 &
$3.4^{+0.2}_{-0.2}$ & 1.6 & 88 \\
Obs12 & 0.376 & $3.8^{+0.1}_{-0.2}$ & 51.5 & $8.3^{+2.5}_{-1.9}$ & 16.6 &
$3.1^{+0.1}_{-0.1}$ & 0.9 & 57 \\
Obs13 & 0.053 & $3.3^{+0.2}_{-0.5}$ & 51.4 & $6.5^{+1.1}_{-0.8}$ & 3.2 &
$3.3^{+0.2}_{-0.2}$ & 1.2 & 69 \\
Obs14 & 0.209 & $3.4^{+0.2}_{-0.3}$ & 51.5 & $6.7^{+0.9}_{-0.4}$ & 3.0 &
$3.7^{+0.2}_{-0.2}$ & 1.2 & 87 \\
Obs15 & 0.768 & $2.7^{+0.5}_{-0.5}$ & 51.2 & $5.4^{+0.4}_{-0.3}$ & 0.8 &
$3.6^{+0.3}_{-0.2}$ & 1.8 & 56 \\
\\
\hline
\end{tabular}
\tablefoot{
\tablefoottext{a}{Two-component isothermal model with fixed abundance ratios
(O/Fe=0.3, Ne/Fe=0.5, Mg/Fe=0.9, and other element abundances equal to
Fe = 0.5 Fe$_{\odot}$).}
\tablefoottext{b}{X-ray luminosity in the 0.2-2.5\,keV band.}
}
\end{center}
\end{table*}

\normalsize

\begin{figure}
\centering
\includegraphics[viewport=1 1 405 360,clip,width=\linewidth]{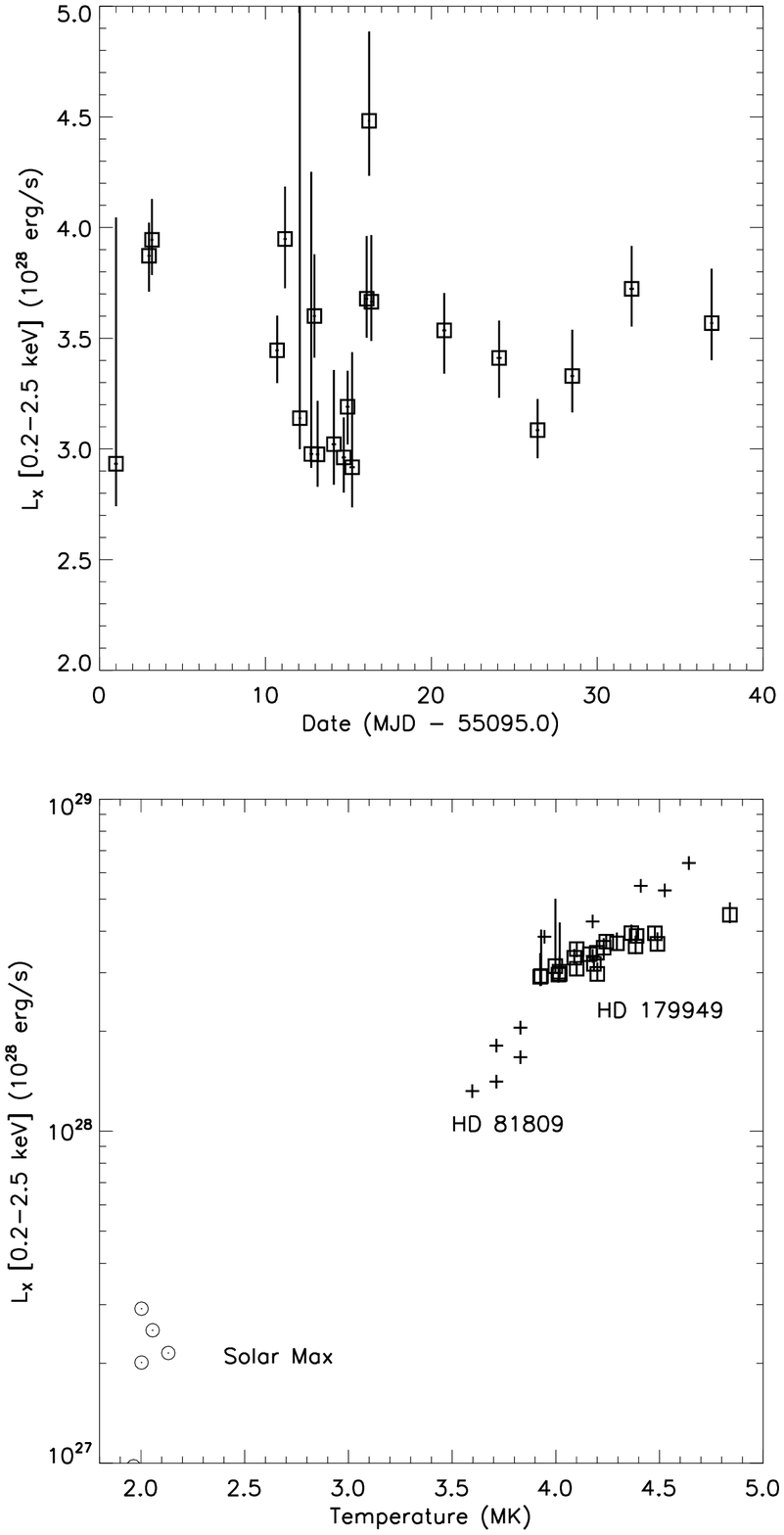}
\caption{
X-ray luminosity vs.\ coronal temperature for our target, \object{HD\,179949},
and two other comparison stars (see text). Note that the
temperature was computed as the mean of the two components, weighted by
their respective plasma emission measures.
}
\label{fig:fitres}
\end{figure}

The cooler component in the 2-T model has a temperature $T_{1} \sim 3-4$~MK, while the hotter component has  $T_{2} \sim 5-9$~MK. The ratio of the plasma emission measures of these two components is in the range $EM_{1}/EM_{2} \sim 1-17$. The best-fitting models allowed us to compute the broad-band (0.2-2.5\,keV)\footnote{For computing the X-ray luminosity we choose the 0.2-2.5 keV band for ease of comparison with similar data taken from the literature. In any case, we have verified that an extension of the bandpass to 10 keV implies a flux increase by less than 1\%, for all our best-fitting models, because most of the emission is associated with the cooler component, and thermal spectra are characterized by an exponential decrease at large photon energies.} source X-ray luminosity and an average coronal temperature weighted by the emission measure. Figure~\ref{fig:fitres} 
 shows a scatter plot of the X-ray luminosity vs.\ the average coronal temperature.

\subsection{Temporal variability of the X-ray luminosity}

We find that the broad-band X-ray luminosity is well correlated with the average coronal temperature, a dependence typical of solar-type stars, and also considering the same star at different observing times. In Fig.~\ref{fig:fitres} we show a comparison of \object{HD\,179949} with measurements of coronal temperature and luminosity for the \lq\lq integrated\rq\rq\ Sun at solar maximum \citep{Orlando2001}, and a number of data points for HD\,81809, a G-type star monitored for several years and showing evidence of a solar-like magnetic cycle both in the chromospheric (\ion{Ca}{ii} H\&K) and in the coronal X-ray emission \citep{Favata2008}. We conclude that most of the observed variability of our target is due to chromospheric and coronal activity, and any effect due to the presence of the planet is difficult to disentangle.

\begin{figure}
\centering
\includegraphics[width=\linewidth]{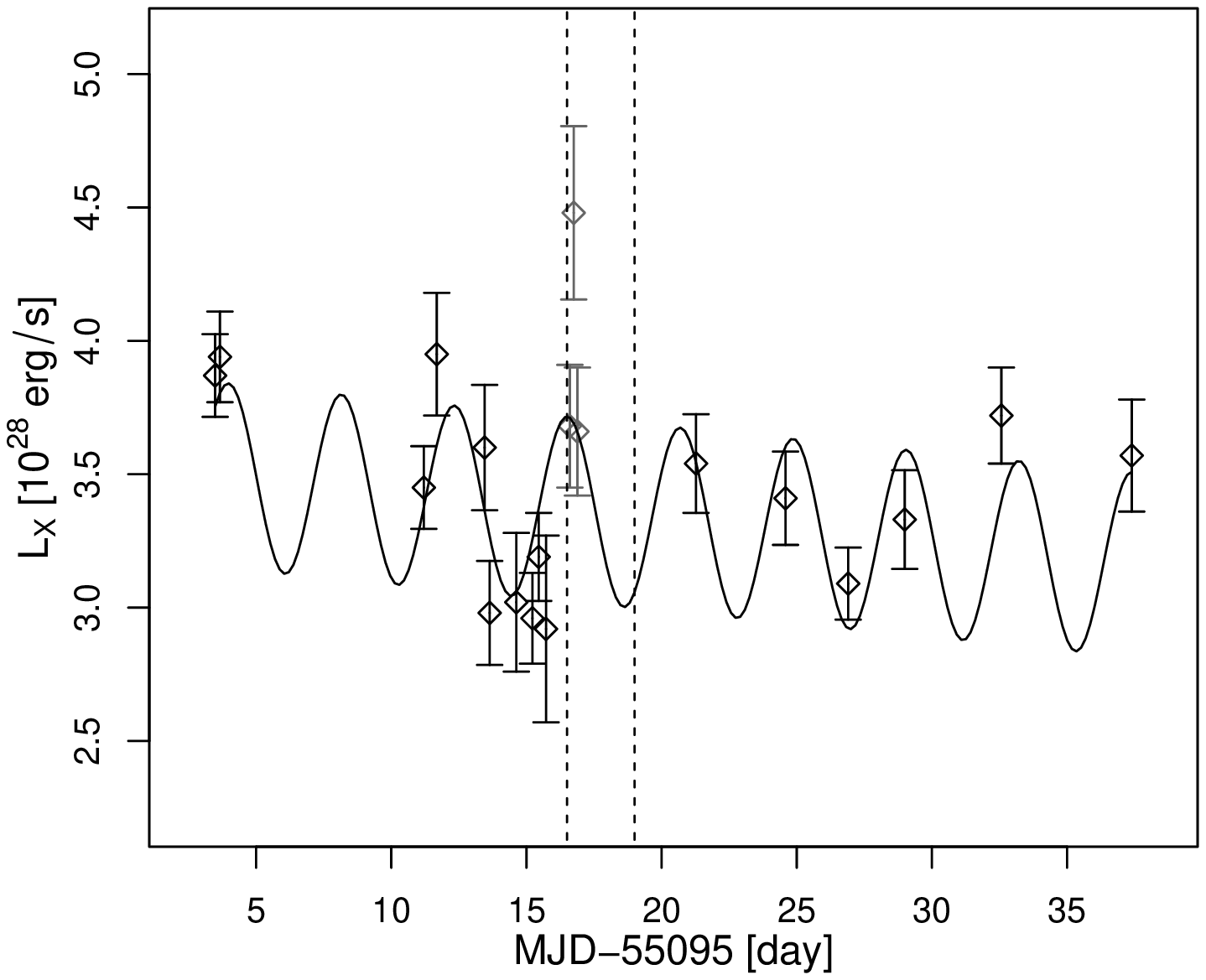}
\includegraphics[width=\linewidth]{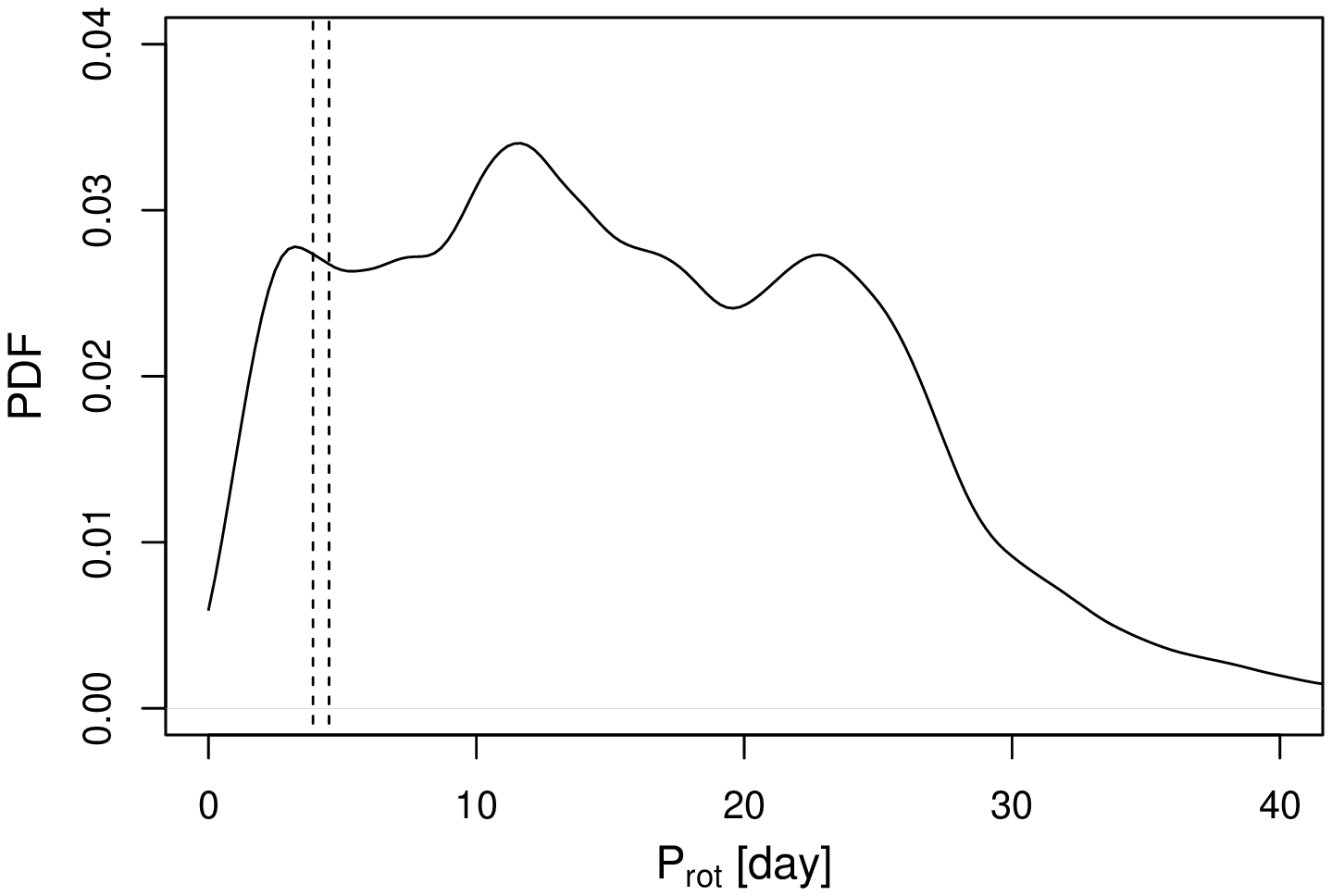}
\caption{\textit{Top panel - }Best fit of the X-ray light curve, using a linear term and one sinusoidal modulation. The vertical dotted lines bracket the time range likely contaminated by the flare activity and excluded in the fitting, as discussed in Sect.~\ref{sec:time}. The fitted period is $P=$4.2$\pm0.1$~days. \textit{Bottom panel - }Probability Distribution Function of the fitted periods recovered after 10,000 time shufflings. The FAP is $\sim$2\%.}\label{fig:fit_X}
\end{figure}

To support this conclusion, in Fig.~\ref{fig:fit_X} we show the X-ray light curve of the data with uncertainties $\lesssim$30\%. Following the approach introduced in Sect.~\ref{sec:time}, and excluding the days with flare activity, we find that this time series is fitted with $\chi_{red}^{2}\sim$2.2 by considering a linear trend and a sinusoidal modulation with $P$=4.2$\pm$0.1~days. The confidence level for this fit is quite low ($p\sim$1.6\%), while the FAP is $\sim 2$\% (Fig.~\ref{fig:fit_X}, bottom panel).

The fitted period is close to half the rotation period at intermediate latitudes \citep{Faresetal12} and can be interpreted with the presence of two active regions co-rotating with the star on opposite hemispheres. Alternatively, it is consistent with the beat period between the stellar equatorial rotation and the planet orbit, but we regard this similarity as a spurious coincidence since we do not find any evidence of a SPI signal with the same periodicity in the chromospheric time series.

Regarding the flaring event observed on Oct 6 2009, it occurred within a few hours from simultaneous du Pont and CFHT observations which provided optical spectra characterized by enhanced emission in the core of the \ion{Ca}{ii} H\&K lines. High chromospheric activity is also indicated by further du Pont and TNG observations on Oct 7 and 8 2009. Taken as a whole, these data suggest that the X-ray flare occurred during a period of high activity on \object{HD\,179949}, which perdured for $\sim 3$~days, i.e.\ one full planetary orbit. In the absence of simultaneous X-ray data, the enhanced \ion{Ca}{ii} signal could have been included blindly in the search for a periodic modulation of the chromospheric signature, thus introducing significant non-periodic features in the analyzed time series. On the other hand, we cannot exclude that this event is also related to SPI.



\section{Discussion and conclusions}

We have conducted a nearly simultaneous optical and X-ray observational campaign on the planetary host star \object{HD\,179949} in September-October 2009, aimed at collecting new data for a better understanding of its activity and the possible magnetic star-planet interaction. Our observational campaign covers $\sim$5 orbital periods and $\sim$2 stellar rotations. The modeling of the residual fluxes in the cores of the \ion{Ca}{ii} H\&K lines shows that the most likely explanation for the observed variability of the chromospheric flux is the rotational modulation  produced by several active regions. The simplest model assumes a modulation with a rotation period of $\sim 11$ days and its first and second harmonics. Such a period corresponds to the high-latitude rotation period of the star as found by \citet{Faresetal12}, who derived its surface differential rotation from the analysis of Zeeman Doppler Imaging maps. The inclination of the stellar rotation axis, as derived from Doppler Imaging, is $\approx 60^{\circ}$, indicating that one of the poles is always in view and suggesting that the amplitude of the modulation produced by equatorial active regions is smaller than the modulation due to intermediate or high latitude regions, owing to the effect of the foreshortening. On the other hand, the modulation of the X-ray coronal flux suggests the presence of two active regions on opposite hemispheres of the star rotating with a period of $\sim 9$~days, that corresponds to low latitudes. These results indicate that the active regions responsible for the chromospheric and coronal variabilities are not co-spatial, as often observed in the Sun. 
In HD~179949, the pressure scale height of the coronal loops ranges from $\approx 1.5 \times 10^{10}$ to $\approx 4.5 \times 10^{10}$~cm, i.e., it is a significant fraction of the stellar radius (17--52\%), implying that the coronal emission is not necessarily confined to levels close to the stellar surface. 

The scenario with a component of the chromospheric and coronal variability being phased with the orbital period of the planet, or its beat period with respect to the rotation period at some stellar latitude, is regarded as unlikely in view of the results of our analysis. When we fit the data excluding the time interval with enhanced activity (marked by an X-ray flare) and fixing the period of the planet, we find a significantly worse fit than assuming only rotational modulation (cf.\ Sect.~\ref{sec:time}). When we keep all the data, we find a secondary periodicity at $\simeq 4.2$ days: this value is close to the second harmonic of the stellar rotation period $P_{\rm rot} \simeq 11$ days, but also to the beat period between the planet orbital period and the stellar rotation period, as expected in the case of a magnetic SPI \citep[cf., e.g., ][]{Lanza12}. However, the quality of the fit is poor and it is not possible to discriminate between the latter two alternatives. 

The possibility that the intrinsic evolution of the active regions hides the SPI signal cannot be excluded, of course. Considering the results in Sect.~\ref{sec:time},  the difference of $\sim 0.5$ days between the secondary period of \textit{Model C} ($P_{sec}$=3.9$\pm$0.1~day) and the orbital period of the planet could be produced by the intrinsic evolution of the active regions on the star or by their shearing owing to the differential rotation. Assuming a simple model in which the amplitude of the SPI signal is modulated in a sinusoidal fashion with an evolutionary period $P_{\rm evol}$, we find that the 3.6-day period is obtained when $P_{\rm evol} \simeq 25$~days, i.e. slightly longer than the extension of our time series and not significantly different from the pole-equator lap time of $\sim 29$~days that  comes from the differential rotation parameters of \citet{Faresetal12}. Unfortunately, it is not possible to prove this conjecture because the number of data points is limited, so that no significant result can be obtained by splitting the time series in two halves and analyzing separately each subseries to search for the effects of the intrinsic spot evolution and/or migration. We may also note the coincidence between the above period difference and the difference between the polar rotation period found by \citet{Faresetal12}, i.e., $10.3 \pm 0.8$~days, and the rotation period of \textit{Model A}, i.e., $10.9 \pm 0.2$~days. However, given the large uncertainties, these values are compatible with each other, and the claim that the active region evolution affects also the determination of the rotation period cannot be supported.  

When comparing our results with those of \citet{Faresetal12}, one should consider that the present time series is significantly longer than the previous one.  Based on the smaller data set consisting only of the ESPaDOnS spectra, \citet{Faresetal12}  modeled the chromospheric variability in the \ion{Ca}{ii} H\&K and H$\alpha$ lines by assuming a model  identical to our \textit{Model C}. It provided a rotation period $P_{1}=8.4$~days and an additional periodicity at $P_{3}=4.9$~days, that was tentatively associated with a beating between the orbital motion and the stellar rotation at some intermediate latitude. However, the present result, based on a longer time series and the removal of the data points affected by flaring, points toward a modulation produced by stellar rotation without any evidence of a planet-induced component.  Since the chromospheric signatures of the star-planet interaction are present in approximately 50\% of the seasons \citep{2008ApJ...676..628S}, our apparently negative result is in line with the expected frequency. Further observations would be valuable to confirm the presence of a signature moving in phase with the orbital motion of the planet, as found in 2001--2002 and 2005--2006 by \citet{2008ApJ...676..628S} and \citet{Gurdemiretal12}. The transition from an \textit{on} to an \textit{off} state of chromospheric SPI could be associated with the variation of the stellar field configuration along an activity cycle. A tentative period of $\approx 4-5$ years could be derived from the above results. However, the lack of an observed signature in 2009, also confirmed by \citet{Poppenhaegeretal11}, indicates that the on-off cycles may not be periodic. As a matter of fact, the numerical simulations by \citet{Cohenetal10,Cohenetal11} show that the appearance of a signature in the stellar atmosphere is associated with the planet's magnetosphere being inside the stellar Alfven surface or at least in contact with it. If the planet is outside the Alfven surface, the strength of the interaction is remarkably reduced and the perturbations produced by the motion of the planetary magnetosphere cannot propagate back to the stellar surface because the stellar wind blows them away. Since the changes in the stellar field intensity and geometry along the activity cycle can modify the radius of the Alfven surface up to a factor of $1.5-3$, we expect a remarkable variation of the SPI signal and even its disappearance during the time intervals in which the stellar Alfven surface retreats inside the planetary orbit.

\begin{acknowledgements}

This work is based on multi-wavelength observations obtained with the {\it XMM-Newton} telescope, the {\it Canada-France-Hawaii Telescope} (CFHT), the Italian {\it Telescopio Nazionale Galileo} (TNG) and the  {\it Ir\'en\'ee du Pont} telescope.

XMM-Newton, an ESA science mission with instruments and contributions directly funded by ESA Member States and the USA (NASA).

CFHT is operated by the National Research Council of Canada, the Institut National des Science de l’Univers of the Centre National de la Recherche Scientifique of France, and the University of Hawaii. The ESPaDOnS data were reduced using the data reduction software Libre-ESpRIT, written by J.-F. Donati from Observatoire Midi-Pyrenees, and provided by CFHT \citep{Donati97}.

The SARG spectrograph on the TNG is operated on the island of La Palma by the {\it Fundacion Galileo Galilei} of the INAF {\it (Istituto Nazionale di Astrofisica)} at the Spanish {\it Observatorio del Roque de los Muchachos} of the {\it Instituto de Astrofisica de Canarias}.

The Ir\'en\'ee du Pont telescope has been in operation at Las Campanas Observatory since 1977. The telescope was a result of a gift in 1970 from Mr.\ and Mrs.\ Crawford H.\ Greenewalt to the Carnegie Institution of Washington, which supplied supplemental funds.

We thank an anonymous referee for carefully reading the manuscript and providing suggestions that significantly improved our paper.
\end{acknowledgements}


\end{document}